\documentclass{JHEP3}  \preprint{}

\usepackage{cite}
\usepackage{epsfig}
\usepackage{subfigure}
\usepackage{color}

\usepackage{graphicx}
\usepackage{inputenc}
\usepackage{xspace}
\inputencoding{latin1}
\usepackage{epstopdf}
\def\br{BR}
\def\tanb{\tan\beta}
\def\gev{~{\rm GeV}}
\def\tauptaum{\tau^+\tau^-}
\def\anti{\overline}

\def\lsim{\mathrel{\raise.3ex\hbox{$<$\kern-.75em\lower1ex\hbox{$\sim$}}}}
\def\gsim{\mathrel{\raise.3ex\hbox{$>$\kern-.75em\lower1ex\hbox{$\sim$}}}}
\def\ifmath#1{\relax\ifmmode #1\else $#1$\fi}

\def\what{\widehat}
\def\vev#1{\langle #1 \rangle}

\title{Reinstating the `no-lose' theorem for NMSSM Higgs discovery at the LHC}
\preprint{MAN/HEP/2007/44 \\ UCD-07-03}
\author{J.~R.~Forshaw$^1$, J.~F.~Gunion$^2$, L.~Hodgkinson$^1$, A.~Papaefstathiou$^{1,3}$ and A.~D.~Pilkington$^1$\\$^1$School of Physics \&
Astronomy, University of Manchester, Manchester M13 9PL, UK.\\
$^2$Department of Physics, University of California, Davis CA 95616-8677, USA. \\ 
$^3$Cavendish Laboratory, University of Cambridge, Cambridge CB3 0HE, UK.\\
\email{forshaw@mail.cern.ch }\email{gunion@physics.ucdavis.edu}}

\abstract{The simplest supersymmetric model that solves the $\mu$
  problem and in which the GUT-scale parameters need not be finely
  tuned in order to predict the correct value of the $Z$ boson mass at
  low scales is the Next-to-Minimal Supersymmetric Standard Model (NMSSM).
  However, in order that fine tuning be absent, the lightest CP-even
  Higgs boson $h$ should have mass $\sim 100\gev$ and SM couplings
  to gauge bosons and fermions.  The only way that this can be
  consistent with LEP limits is if $h$ decays primarily via $h \to a a
  \to \tauptaum\tauptaum$ or $4j$ but not $4b$, where $a$ is the
  lighter of the two pseudo-scalar Higgses that are present in the
  NMSSM. Interestingly, $m_a<2m_b$ is natural in the NMSSM
  with $m_a> 2m_\tau$ somewhat preferred.  Thus, $h\to
  \tauptaum\tauptaum$ becomes a key mode of interest. Meanwhile, all
  other Higgs bosons of the NMSSM are typically quite heavy.
  Detection of any of the NMSSM Higgs bosons at the LHC in this
  preferred scenario will be very challenging using conventional
  channels. In this paper, we demonstrate that the $h\to
  aa\to\tauptaum\tauptaum$ decay chain should be visible if the Higgs
  is produced in the process $pp \to p+h+p$ with the final state
  protons being measured using suitably installed forward detectors.
  Moreover, we show that the mass of both the $h$ and the $a$ can be
  determined on an event-by-event basis.}

\begin{document}

\section {Introduction}

The Next-to-Minimal Supersymmetric Standard Model (NMSSM) extends the
MSSM by the introduction of a singlet superfield, $\what{S}$. To do so
provides the possibility to solve \cite{dermisekgunion} the fine
tuning problem that arises in the case of the MSSM and it provides a
natural solution to the $\mu$-problem via the $\lambda \what S\what
H_u\what H_d$ superpotential term when the scalar component of $\what
S$ acquires a vev ($\mu_{eff}=\lambda \vev{S}$). The Higgs sector of
the NMSSM contains three CP-even and two CP-odd neutral Higgs bosons,
and a charged Higgs boson. The lightest CP-even Higgs boson is denoted
by $h$ and the lightest CP-odd Higgs boson is denoted by $a$.

In \cite{Gunion:1996fb} and \cite{Gunion2}, a partial `no-lose'
theorem for NMSSM Higgs discovery was established. The theorem states
that the LHC would be able to detect at least one of the NMSSM Higgs
bosons in conventional modes provided Higgs-to-Higgs decays are not
important.  However, Higgs-to-Higgs decays can dominate over other
decays, especially for a light Higgs boson, because of the small size
of fermionic-Higgs Yukawa couplings.  This was first established in
\cite{Ellis:1988er} with early subsequent discussions in
\cite{Gunion:1996fb} and \cite{Dobrescu:2000jt} followed by studies
such as that of \cite{Gunion3} and \cite{Gunion1}.  In fact, in
\cite{dermisekgunion} it was found that the part of NMSSM parameter
space where the model has no fine-tuning problem (see later) is such
that Higgs-to-Higgs decays must be taken into account and do dramatically
affect LHC Higgs boson search strategies. In this part of parameter
space, the lightest CP-even Higgs boson has mass $m_h\sim 90-100\gev$
and SM-like gauge and fermion couplings. However, the width for
$h$ to decay to a pair of the lightest CP-odd Higgs bosons is much
larger than the width for $h$ decay to $b\anti b$, $\Gamma(h\to aa)\gg
\Gamma(h\to b\anti b)$, as is easily possible given the very small
$b\anti b h$ SM-like Yukawa coupling.  The $a$ has no tree-level
coupling to $ZZ$ and would have escaped LEP searches. In addition, the
$a$ is very weakly produced in hadron collisions and unlikely to be
observable at the LHC.  The $h$ with $m_h\sim 100\gev$ would also have
escaped detection at LEP even though its mass is below $114\gev$.  The
exact lower bound on $m_h$ with dominant $h\to aa$ decays depends upon
how the light CP-odd $a$ itself decays. If $a\to b\anti b$, then the
$h\to b\anti b$ and $h\to b\anti b b\anti b$ channels must be
combined.  The lower limit on $m_h$ depends somewhat on the relative
branching ratios for these two channels but is typically of order
$110\gev$ when the latter is dominant~\cite{bechtlepc}. As a result,
the non-fine-tuned scenarios (which require $m_h\sim 100\gev$) only
survive LEP limits if $m_a<2m_b$. If $a\to \tauptaum$, then the lower
bound on $m_h$ is close to $89\gev$~\cite{Schael:2006cr}. If $a\to 2j$, then
the only applicable lower bound is that where $h$ is assumed to decay
hadronically but not necessarily into two jets.  This lower bound is
approximately $82\gev$~\cite{Abbiendi:2002qp}. Thus, scenarios with
$m_h\sim 100\gev$ and dominant $h\to aa\to
\tauptaum~\mbox{or}~4j$ decays are not ruled out by any LEP constraints.

Although the $m_h\sim 100\gev$ with dominant $h\to aa$ decay scenario
would a priori seem somewhat unlikely, in fact it is preferred.
In order to avoid fine-tuning ({\it i.e.} to avoid tuning GUT scale soft-SUSY-breaking parameters 
so as to obtain the correct value of $m_Z$) the superpartners (in particular stops) should not be very heavy.
In this case, the model automatically predicts that the lightest
CP-even Higgs boson should have $m_{h}\sim 100\gev$ for
$\tanb>10$, decreasing slowly to $\sim 90\gev$ for low
$\tanb$~\cite{Dermisek:2005gg}. In other words, if the top squarks are
light then the tree-level mass (which is predicted to be $\sim m_Z$ for
larger $\tanb$ values) is only modestly corrected by the top+stop
loops.  Further, in the NMSSM there is a $U(1)_R$ symmetry that, if
preserved at low scales, would lead to the lightest CP-odd scalar
being massless.  If this symmetry is present at the GUT scale,
small soft-SUSY-breaking terms that violate the $U(1)_R$ symmetry are
automatically generated at low scales by renormalization group
evolution and one most naturally obtains a light
$a$~\cite{Dermisek:2006wr}. In fact, as we have already noted, if
$m_{a}>2m_b$ so that $h\to aa\to 4b$, then LEP excludes
$m_{h}<110\gev$.  Thus, for the non-fine-tuned $m_{h}\sim 100\gev$
case, we are essentially uniquely led to the $h\to aa\to
\tauptaum\tauptaum~\mbox{or}~4j$ scenarios.  Further, the study
of~\cite{Dermisek:2005gg} finds that $m_a>2m_\tau$ (but $m_a<2m_b$)
is possible without any tuning of the relevant GUT-scale
soft-SUSY-breaking parameters whereas $m_a<2m_\tau$ can require some
tuning (see also the discussion below).  This leads us to a preference for the $h\to aa\to
\tauptaum\tauptaum$ scenario.  Finally, it is important to note that
$m_h\sim 90-100\gev$ has two additional attractive features: (a)
the predicted value of $BR(h\to aa) >0.75$ leads to a reduced value for
$BR(h\to b\anti b)$ that is typically such as to explain the $2.3\sigma$
event excess in the $e^+e^-\to Z b\anti b$ final state for $M_{b\anti
  b}\sim 98\gev$~\cite{dermisekgunion}; (b) such $m_h$ provides
the best description of precision electroweak data. Thus, the scenario
envisioned has excellent phenomenological and theoretical properties.

\begin{figure}[t]
  \centering \mbox{
    \includegraphics[width=0.50\textwidth]{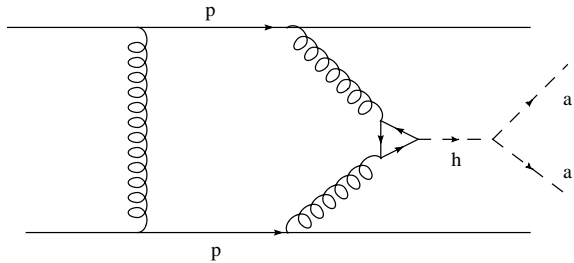}\qquad }
\caption{Central exclusive NMSSM Higgs production at the LHC.}
\label{fig:higgscep}
\end{figure}

Our strategy will be to utilize the central exclusive production (CEP) process, $pp \to p+h+p$ with $h \to aa$, which is illustrated in Figure~\ref{fig:higgscep}. The outgoing protons carry almost all of the incoming beam energy and it is understood that they should be detected in purpose built low-angle detectors \cite{Cox,fp420}.  These high-resolution sub-detectors would permit the four-momentum of the central system to be determined accurately and it is our purpose to exploit this capability to eliminate backgrounds and extract the masses of both the $h$ and $a$ on an event-by-event basis. We note that the possibility to use CEP to investigate new physics (especially supersymmetric new physics) has been much studied over the past few years, see \cite{prospects,SUSY,uncertainties,Bussey:2006vx}.

\begin{figure}[h!]
\begin{center}
\includegraphics[width=0.5\textwidth,angle=90]{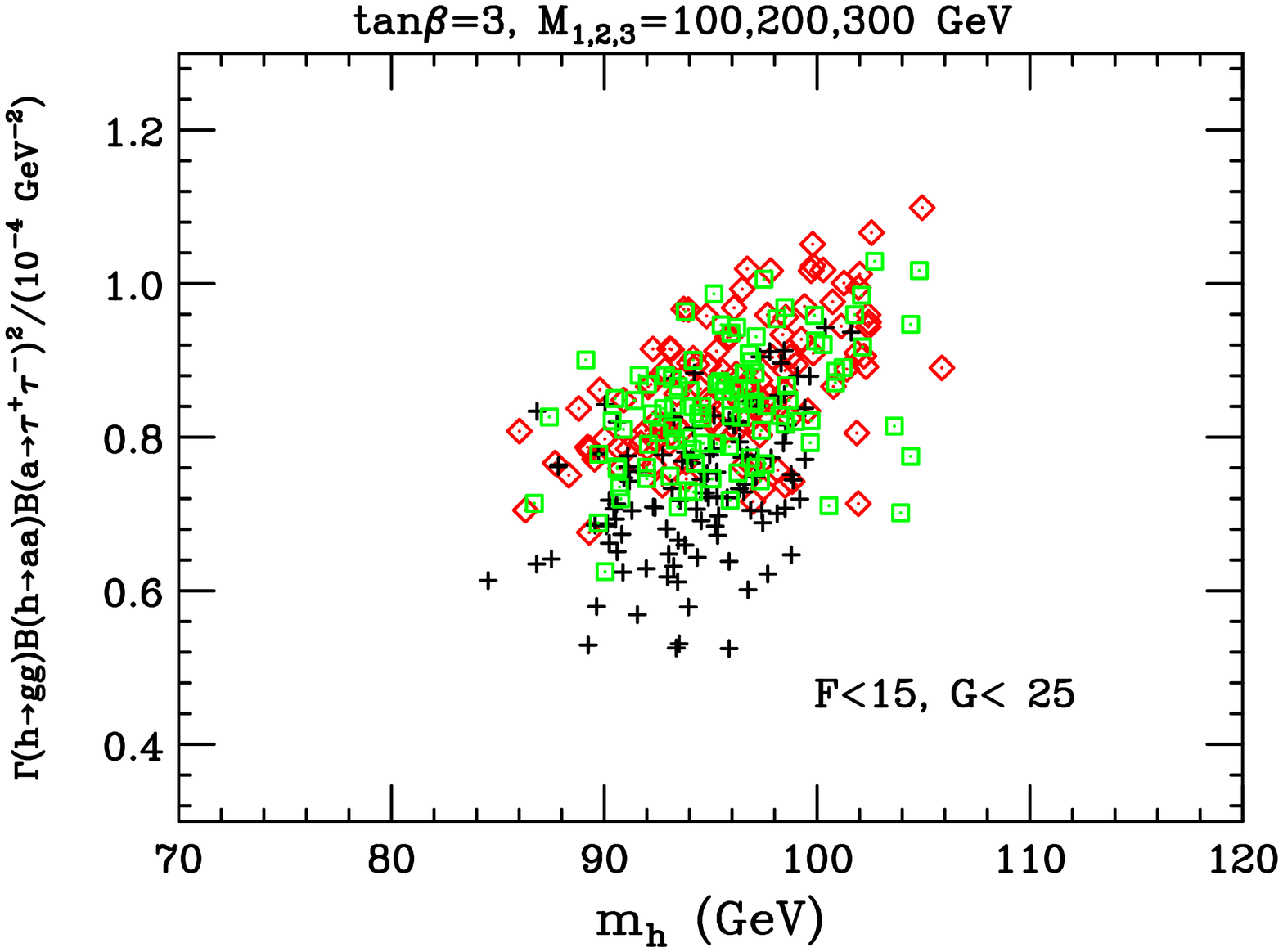}\\
\includegraphics[width=0.5\textwidth,angle=90]{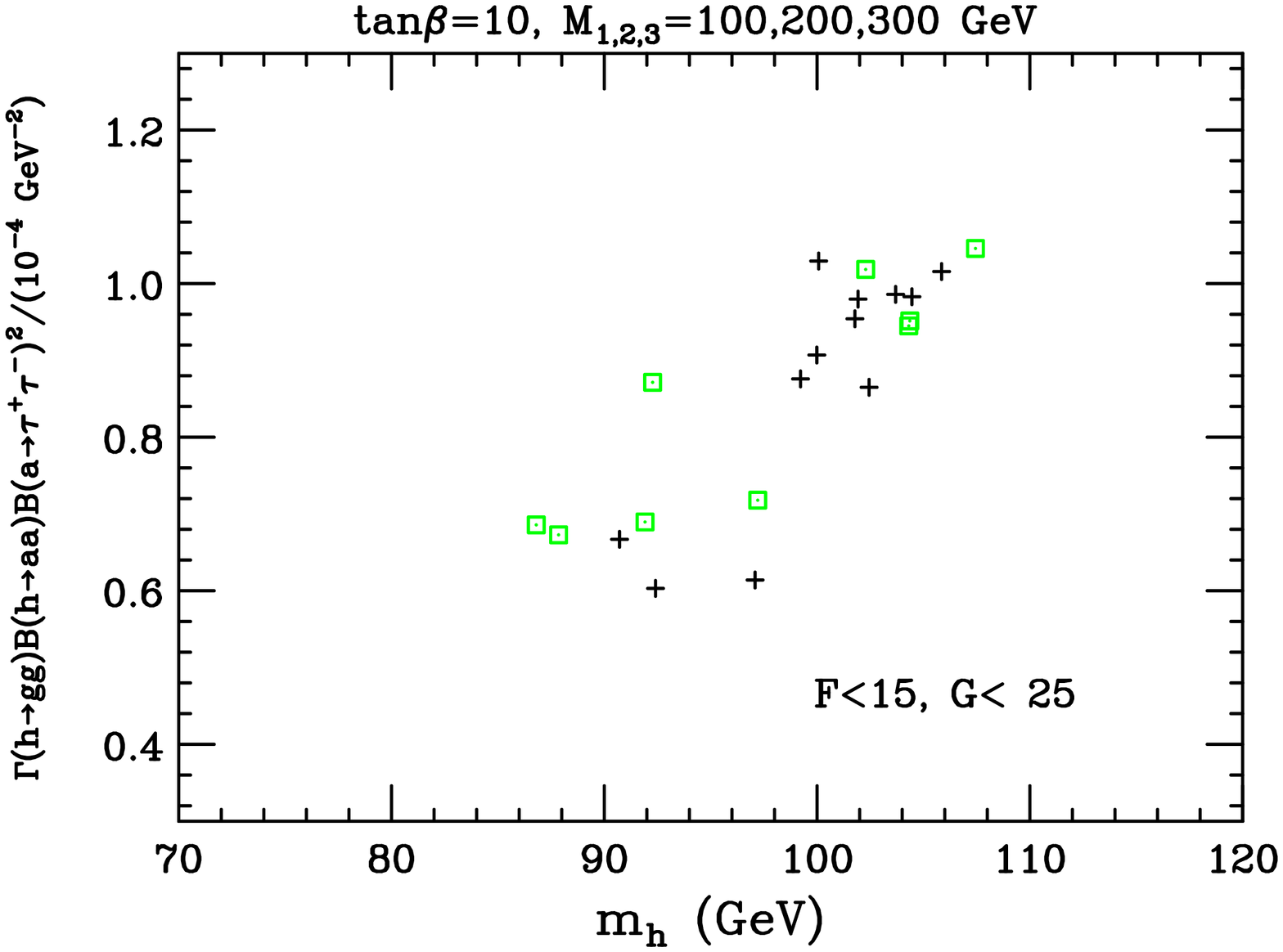}
\caption{$\Gamma_{\mathrm{eff}}$ in units of $10^{-4}\gev^2$ versus $m_h$ for
  $\tanb=3$ (top) and $\tanb=10$ (bottom). Point coding: (red) diamonds =
  $2m_\tau<m_a<7.5\gev$; (green) squares = $7.5\gev\leq m_a <8.8\gev$;
  (black) pluses = $8.8\gev\leq m_a<2m_b$.}
\label{fig:scatter}
\end{center}
\end{figure}

The $pp\to  p+4\tau+p$ event rate is proportional to
\begin{equation}
\frac{\Gamma_{\mathrm{eff}}}{m_h^3} \equiv \frac{\Gamma(h\to gg)}{m_h^3}\br(h\to aa)\br(a\to \tauptaum)^2 \; 
\label{eq:gghxsec}
\end{equation}
and to a $m_h$-dependent luminosity function, which characterises the flux of fusing gluons.
For those points with minimal electroweak fine-tuning ($F < 15$ in the notation of
\cite{dermisekgunion,Dermisek:2005gg}\footnote{$F < 15$ corresponds to fine-tuning of
GUT-scale parameters no worse than 6.5\%.}) there is a factor $\sim 3$ variation in the value of $\Gamma_{\mathrm{eff}}$.  If one also imposes  that the ``light-$a$-fine-tuning'' parameter $G$ (defined in \cite{Dermisek:2006wr}) obeys $G < 25$ in order
to achieve $m_a<2m_b$ and $\br(h\to aa)>0.7$ (as required to
escape LEP limits)\footnote{$G < 25$  corresponds to no worse than
4\% tuning of the $A_\kappa$ and $A_\lambda$ parameters of the NMSSM.}, then the variation  of $\Gamma_{\mathrm{eff}}$ is reduced to a factor $\sim 2$, with the precise amount of variation depending somewhat on $m_h$, $\tanb$ and $m_a$.  
The dependence of $\Gamma_{\mathrm{eff}}$ upon $m_h$
is presented in Figure~\ref{fig:scatter}, for two different values of $\tanb$. The points
correspond to points in parameter space which have $F<15$ and $G<25$ and
they are coded according to $m_a$. Scatter plots at these two values of $\tanb$ cover the most natural region of NMSSM parameter space, since for larger value of $\tanb$ it is not possible to keep both $F$ and $G$ sufficiently small.  We note that $m_a<2m_\tau$ generally leads to $G>50$, for all values of $\tanb$, and thus the four-tau decay channel is always most natural.  The second part of the production cross-section is the gluon luminosity function and it falls by a factor $\sim 2$ in going from $m_h=90\gev$ to $m_h=110\gev$ \cite{prospects}. Including the explicit $1/m_h^3$ dependence of 
(\ref{eq:gghxsec}), the production cross-section therefore falls by a factor $\sim 3$ as $m_h$ increases from 90 GeV to 110 GeV at fixed $\Gamma_{\mathrm{eff}}$.

In the present paper we focus our attention on a scenario for which
the scalar Higgs, $h$, has mass 92.9 GeV, and the pseudo-scalar Higgs,
$a$, has mass 9.73 GeV. This scenario can be achieved with $F<15$ and $G<25$ for
either $\tanb=3$ or $\tanb=10$. For our study, we employ a value of
$\Gamma_{\mathrm{eff}}=0.50\times 10^{-4}\gev^2$, {\it i.e.} the lowest value
that can be achieved according to the scatter plots of
Figure~\ref{fig:scatter}.  The value of $m_h$ chosen is in the middle of the preferred range of
values appearing in Figure~\ref{fig:scatter}, implying a mid-range value
for the gluon luminosity.  According to the scatter plots of
Figure~\ref{fig:scatter}, there is a trend whereby larger $m_h$ is
associated with larger $\Gamma_{\mathrm{eff}}$, so that this
particular choice of parameters gives a net production rate for the $4\tau$
final state that is not far from minimal.  For example, for
$m_h=93\gev$ one finds in Figure~\ref{fig:scatter} values of
$\Gamma_{\mathrm{eff}}$ as high as $0.9\times 10^{-4}\gev^2$ for $m_a>8.8\gev$.
The particular set of $m_h$, $m_a$ and $\Gamma_{\mathrm{eff}}$ values we employ
in this study corresponds to Scenario 1 of \cite{Gunion1} which used the
version of NMHDECAY \cite{nmhdecay} available at that
time.\footnote{Small improvements in NMHDECAY over the intervening years
  imply that the output $m_h$, $m_a$ and $\Gamma_{\mathrm{eff}}$ values used here are
  slightly changed for the same input parameters used in \cite{Gunion1}.}  The $h\rightarrow aa$ decay occurs with a branching ratio of 92\% and each $a$ decays to
$\tau^{+}\tau^{-}$ with a branching ratio of 81\%.  This implies that
$h$ detection in conventional modes such as the $\gamma\gamma$ final
state or the $\tauptaum$ final state will be highly problematical
and a new detection mode such as the one we study here should be employed.

The remainder of this paper is organized as follows. In the next
section we briefly review the proposal to install forward detectors
420~m either side of the interaction point at ATLAS and/or CMS. In
Section \ref{simulation}, we discuss the simulation of the signal and
relevant backgrounds. In Section \ref{extraction}, our attention turns
to the task of extracting the signal from the variety of backgrounds.
We propose an analysis which is heavily track based in order to
minimize the effect of pile-up.  We then include
an estimation of the combinatorial (or overlap) background caused by
pile-up of multiple $pp$ interactions. We take account of detector
efficiencies, discuss different triggering options and assess the
significance of the Higgs signal for various possible machine
luminosities and detector parameters. In Section
\ref{sec:conclusions}, 
we present our conclusions.
 
\section{The FP420 project} \label{fp420s}

FP420 is an R\&D project with the aim of installing forward proton detectors in the high dispersion region 420~m either side of the interaction point at ATLAS and/or CMS \cite{fp420}. The detectors would measure protons from  CEP, which have lost energy during the interaction and are bent out of the beam by the LHC magnets. There are three issues regarding FP420 which are relevant to this analysis. Firstly, if the protons are tagged, the longitudinal momentum loss, $\xi$, of the protons can be measured to an accuracy of $1-3\%$ \cite{Bussey:2006vx}. This translates, in this paper, into a measurement of the scalar Higgs mass accurate to 2.1~GeV per event. Secondly, the acceptance of FP420 for a central mass of 93~GeV is 28\% for detectors 5~mm from the beam and 38\% for detectors 3~mm from the beam.\footnote{This is in the context of a proton beam for which $\sigma_{\mathrm{beam}}=0.25$~mm. There is also a constant dead region of 0.5~mm to consider at the edge of the proton detectors.} In this analysis we choose the detectors to be 3~mm from the beam, which corresponds to an approximate acceptance at IP1 (ATLAS) given by  
\begin{eqnarray*}
0.002 & \leq  \xi_1 \leq & 0.018 \; , \\
0.0015 & \leq  \xi_2 \leq & 0.014
\end{eqnarray*}
where the asymmetry arises from the LHC beam optics.
Finally, the forward detectors will be capable of measuring the time-of-flight (TOF) of each proton from the interaction point to an accuracy of 10~ps. The difference in the TOF measurements translates into a measurement of the primary vertex position to an accuracy of 2.1~mm, provided the reference timing has negligible jitter. Comparing the vertex implied by FP420 to the vertex measured in the central detector is very useful for rejecting backgrounds; a feature we exploit in Section~\ref{simulation}. In Section \ref{signif}, we present results also for the case pertaining should the 10~ps timing be improved to 2~ps. 

We note that there is also the possibility to use proton detectors installed at 220~m from the IP \cite{Royon:2007ah,Albrow:2006xt}. Such detectors are potentially very valuable since they would be close enough to the interaction point to be included in the level one trigger at ATLAS and CMS, which have latencies of 2.5~$\mu$s and 3.5~$\mu$s respectively. Unfortunately, they do not add much to this analysis due mainly to the low Higgs masses of interest. At CMS, the acceptance for asymmetric tagged protons, i.e. one proton measured at 220~m and one at 420~m, would be at best 5\% for a central mass of 90~GeV \cite{fp420new}. At ATLAS, the corresponding acceptance for asymmetric events could be as large as 20\% if the active edges of the 220~m (400~m) detectors are placed at 2~mm (3~mm) from the beam. However, in this case it has recently been appreciated that there would be an attenuation of the acceptance for symmetric events (i.e. those which use only the 400~m detectors) due to protons that would have been detected at 400~m hitting dead material at the edge of the detectors at 220~m. The net effect is that the combined acceptance at ATLAS would increase by just 5\% after including detectors at 220~m \cite{fp420new}. Furthermore, the OLAP background is harder to estimate and likely to be bigger for asymmetric events because of the possible acceptance of non-diffractively produced protons in the forward detectors \cite{Cox:2007sw}. Finally, the mass resolution of the central system would be greater than the 2.1~GeV which one obtains using 400~m detectors alone. This smears the signal out over a larger region and reduces the significance. For these reasons, the impact of 220~m detectors is likely to be very modest and we do not consider them in the remainder of this paper.

\section{Simulating the signal and backgrounds}\label{simulation}

We have used the ExHuME Monte Carlo event generator (v1.3.4) \cite{Monk:2005ji}, modified in order to simulate NMSSM scalar Higgs production and the decay $h \to aa$.  ExHuME faithfully implements the theoretical approach to CEP detailed in \cite{prospects,KMR2000}.\footnote{Including an overall gap survival factor of 3\% \cite{Khoze:2006uj}.} 
The $a\rightarrow \tau^+ \tau^-$ decays are enforced using the PYTHIA event generator \cite{Sjostrand:2001yu} and PYTHIA is also used for the subsequent $\tau$ decays. For triggering purposes, we consider only those events which contain at least one muon in the final state.\footnote{For both signal and background processes that are simulated using ExHuME we use the CTEQ6M parton distribution functions \cite{Pumplin:2002vw}.}

As the pseudoscalar Higgses are heavily boosted, the signal consists of two back-to-back pairs of taus and we therefore focus our attention on those backgrounds which have similar topology, {\it i.e.} those which look like two jets in the central detector. The central exclusive backgrounds are generated using ExHuME; we simulate $pp\to p+gg+p$ and $pp\to p+b\bar{b}+p$. We do not simulate central exclusive light quark production because these backgrounds are suppressed relative to $b\bar{b}$ by a factor of $m_q^2 / m_b^2$ \cite{KMR2000}. Hence, if we can show that the $b\bar{b}$ background is small, then we show that all di-quark backgrounds are small.

A comment on the theoretical uncertainties associated with our predictions for CEP (signal and background)  is in order, not least because our signal cross-section is small. Studies in \cite{uncertainties,Forshaw:2005qp} have considered these uncertainties in detail and they indicate a factor 2 to 3 uncertainty in the total rate. The main uncertainties arise from the lack of a detailed understanding of the luminosity of the fusing gluons (which is computed using unintegrated gluon densities) and from the difficulty to compute the gap survival factor. The theory has however met with some recent quantitative success: the measurement by CDF of central exclusive dijet production \cite{Aaltonen:2007hs} is in accord with the predictions of ExHuME. Also, it has recently been emphasised that data collected in the early period of LHC running can be expected to shed further light on the theory \cite{Khoze:2008cx}. 

We also consider central particle production arising as a result of `pomeron fusion' (which we usually refer to as `DPE'). In such events, the leading central jets are accompanied by other particle production. We use POMWIG \cite{POMWIG} to simulate this source of background.  POMWIG is able to produce events of the type $pp \to p+jjX+p$ and the version we use \cite{pombeta} incorporates the latest diffractive parton distribution functions from the H1 experiment at HERA \cite{H1}.  Due to a limitation in creating a large enough sample of inclusive dijet events, we concentrate on $b\bar{b}X$  production.  The final $jjX$ background is obtained by re-scaling the number of $b\bar{b}X$ events by the ratio of the $jjX$ to $b\bar{b}X$ cross-sections. Since our analysis is more efficient at eliminating light quark and gluon jets, this procedure should lead us to a conservative estimate of the pomeron fusion background. We discuss the re-scaling in more detail in Section \ref{sec:rescale}. Of course the H1 parton densities come with some uncertainty (see \cite{H1}) and this translates into a modest uncertainty in our prediction of the background from this source provided we assume that the diffractive partons are universal (i.e. that soft re-scattering effects factorize into the gap survival factor).

In addition to these direct backgrounds, at sufficiently high luminosity it becomes necessary to consider the background from what we shall call overlap processes (OLAP). Specifically, we consider the possibility of a threefold coincidence of two single diffractive $pp \to p+X$ events with a generic $pp \to X$ inelastic process. This source of background may be important whenever there is a large number of $pp$ interactions in a bunch crossing. The inclusive QCD events $pp \to X$ are generated using PYTHIA, with the `ATLAS tune' to Tevatron data \cite{jimmytune,Field:2006gq}. For practical purposes, we actually generate $pp \to b \bar{b}+X$ inclusive scattering and re-scale the final results to match the total $2\to2$ scattering cross-section; for details see Section \ref{sec:rescale}.\footnote{We generate $2\to2$ scattering processes with $p_{Tmin} = 4$~GeV which is sufficiently small given that we also require a muon with $p_T>6$~GeV.}
The forward protons (from single diffraction) are then added to the
event using the prescription presented in \cite{Cox:2007sw}, which
also allows us to estimate the probability of the threefold
coincidence as a function of instantaneous luminosity. The two protons
detected by the 420~m detectors do not originate from the same vertex
as the primary scatter which produces the muon and this can be
exploited to reduce the OLAP background. In particular, the $pp$
scattering vertices at ATLAS should be distributed in $z$ by a
Gaussian of width 4.45~cm and in time by a Gaussian of width 0.18~ns
\cite{White:2007ha}. According to the results presented in
\cite{Cox:2007sw}, a rejection factor of 18 (15) should be obtained at
low (high) luminosity running after demanding that the vertex
reconstructed from the forward protons' time-of-flight is required to
fall within $\pm4.2$~mm of the primary vertex.

We do not consider OLAP backgrounds from twofold coincidences. It was shown in \cite{Cox:2007sw} that the largest twofold background --- a coincidence between a soft central diffractive scattering and an inclusive QCD scattering --- was at least a factor of $\sim5$ smaller than the threefold coincidences. This background would be additionally rejected in this analysis as a result of the charged track cut, which is introduced in the next section.

Finally, we consider the pure QED backgrounds: $pp \to p + \tau^+ \tau^- l^+ l^- + p$ (where $l$ is any charged lepton) which is simulated using MADGRAPH \cite{madgraph}. We ignore the effect of the proton electromagnetic form factor ({\it i.e.} we assume pointlike protons) in the simulation but have checked that its effect is not very important in the kinematic regime we consider. The final state tau particles are again decayed using the PYTHIA event generator. Note that we only consider QED backgrounds which include at least one tau pair: non-tau backgrounds can always be eliminated by requiring there to be some missing momentum in the central system ({\it i.e.} using cuts on the $f_i$ variables which we discuss in Section \ref{sec:masses}).

To improve generator efficiency and restrict ourselves to a broad
region of interest, we impose an initial condition that the mass of
the central system lies in the range 70~GeV$\;< M <110$~GeV. All final
state particle four-momenta are smeared according to the relevant
detector component resolution~\cite{atlastdr2} with the outgoing
proton momenta smeared by the amount given in \cite{Bussey:2006vx}.
When appropriate, we also simulate the effects of pile-up by
superimposing additional inelastic $pp$ collisions simulated using
PYTHIA on top of both signal and background events.

\section{Extracting the signal}\label{extraction}

Our analysis is specifically designed to minimize the impact of extra particles in the central detector which arise from multiple proton-proton interactions in each bunch crossing (pile-up). In particular, we propose to utilize mainly tracking information rather than information from the calorimeters. Good vertex resolution for charged tracks  means that the pile-up events rarely contaminate the signal with spurious charged tracks (we quantify this later) and thus we can select events according to the number of charged tracks. We shall select events containing either 4 or 6 charged tracks and at least one muon with $p_T > 6$ GeV. We comment on the effect of a higher muon $p_T$ threshold in Section \ref{trigger}. In this way we collect all signal events in which one tau decays to a muon whilst of the three remaining taus at most one of them is allowed to decay to three charged particles. Without a transverse momentum requirement for the muon, we would expect 47\% of the signal to have one or more muons in the final state. With the muon $p_T$ threshold imposed, we find that 25\% of the signal is available for further analysis.

\subsection{Event selection}\label{analysis}

Before any cuts, the signal is predicted to have a cross-section of
4.8 fb after accounting for the branching ratios for the decay to the
$4\tau$ final state. However for triggering purposes we are forced to
start from a sample of events (signal and background) containing at
least one muon with $p_T > 6$ GeV. We also select only those events in
which the forward protons fall within the acceptance of the 420~m
detectors. The cross-sections after only these cuts are presented in the
first line of Table \ref{tb:sigmas}. In the case of the overlap
background (OLAP), the cross-section is that computed assuming a luminosity of $10^{34}$~cm$^{-2}$~s$^{-1}$ and includes the factor of 15 suppression which we discussed in the previous section and which represents the background rejection from time-of-flight vertexing. We remind the reader that the OLAP background is luminosity dependent and its cross-section scales approximately as ${\cal L}^2$; it is therefore negligible at low luminosities. In this section, we ignore the possibility that pile-up can introduce additional charged tracks but we shall return to justify this in due course.

\begin{table}[ptb]
\begin{center}
\begin{tabular}
{|c|c|c|c|c|c|c|c|} \hline
 &  \multicolumn{3}{c|}{CEP} & DPE & OLAP &\multicolumn{2}{c|}{QED} \\ \hline
Cut & $H$ & $b\bar{b}$ &$gg$ & $b\bar{b}$& $b\bar{b}$ & $4\tau$ & $2\tau$~$2l$ \\ \hline
$p_{T0}^\mu$, $\xi_1$, $\xi_2$, $M$     & 0.442 & 25.14 & 1.51$\times10^{3}$ & 1.29$\times10^{3}$& 1.74$\times10^{6}$ & 0.014 & 0.467 \\ \hline
$N_{\mathrm{ch}} =$ 4 or 6 & 0.226 & 1.59 & 28.84 & 1.58$\times10^{2}$ & 1.44$\times10^{4}$& 0.003 & 0.056\\ \hline
$Q_h = 0$ & 0.198 & 0.207 & 3.77 & 18.69 & 1.29$\times10^{3}$ & 5$\times10^{-4}$& 0.010 \\ \hline
Topology & 0.143 & 0.036 & 0.432 & 0.209 & 1.84 & - & $<$0.001 \\ \hline
$p_T^{\mu}$, isolation & 0.083 & 0.001 & 0.008 & 0.003 & 0.082 & - & - \\ \hline
$p_T^{1, \not{\mu}}$ & 0.071  & 5$\times$10$^{-4}$ & 0.004 & 0.002 & 0.007  & -& - \\ \hline
$m_a > 2 m_\tau$ & 0.066 & 2$\times$10$^{-4}$ & 0.001 & 0.001 & 0.005 & - & - \\ \hline
\end{tabular}
\end{center}
\caption{The table of cross-sections for the signal and backgrounds. All cross-sections are in femtobarns and the cuts are explained in the text. The overlap (OLAP) background is computed at a luminosity of $10^{34}$~cm$^{-2}$~s$^{-1}$.}
\label{tb:sigmas}
\end{table}

\begin{figure}[t]
\centering
\mbox{
  \subfigure[]{\includegraphics[width=0.5\textwidth]{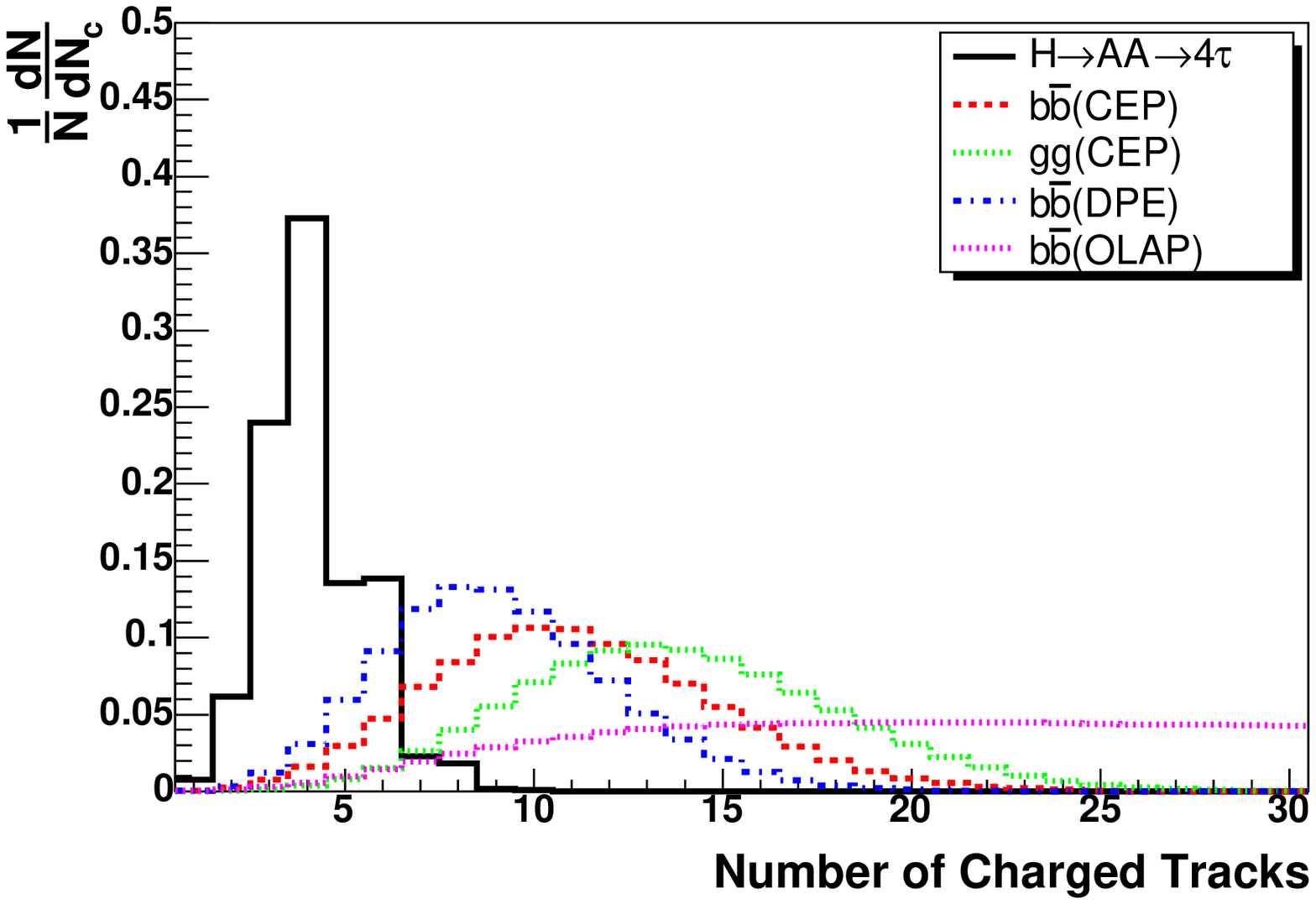}} 
   \subfigure[]{\includegraphics[width=0.5\textwidth]{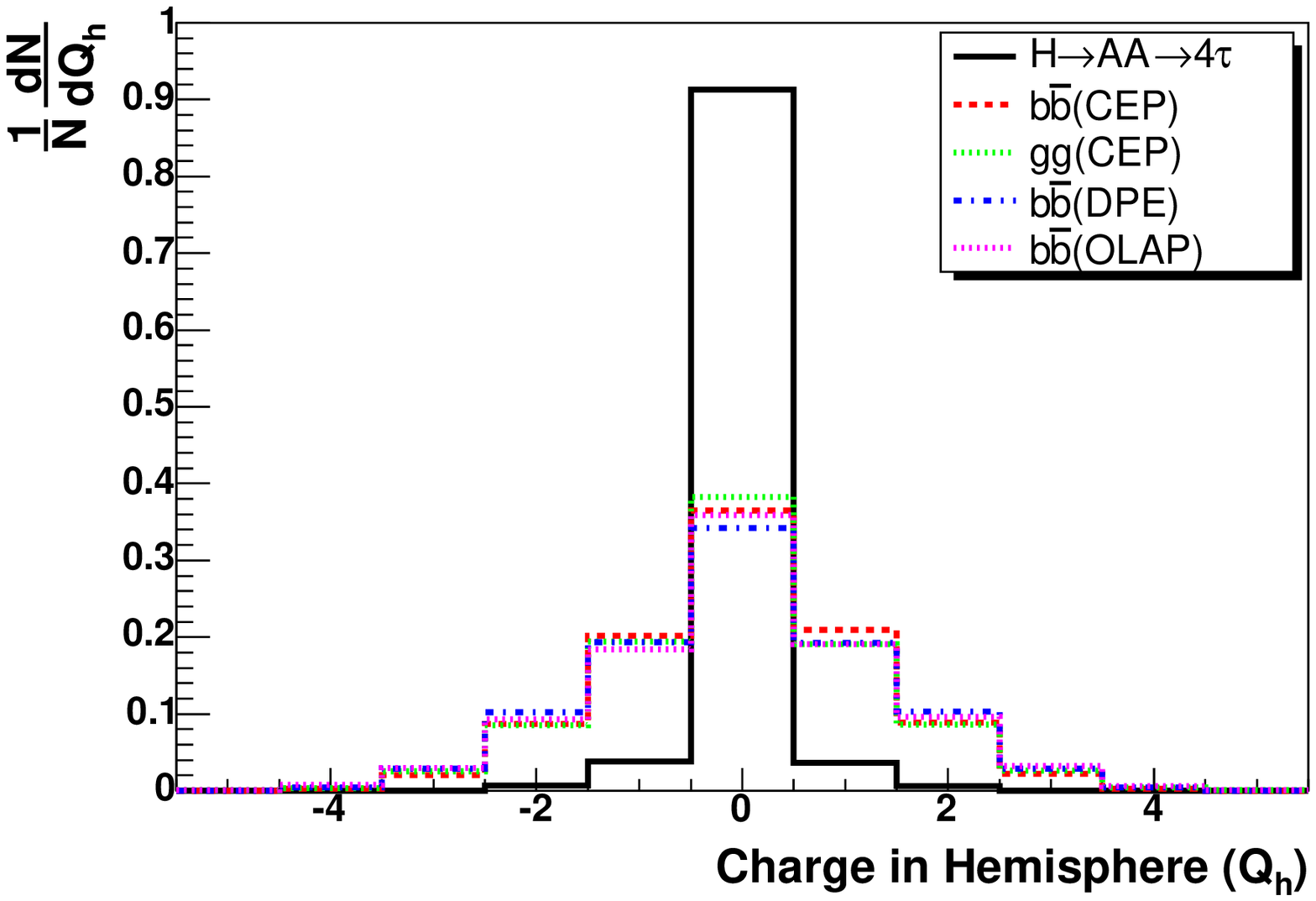}}
}
\caption{(a) The expected number of charged tracks reconstructed by the ATLAS inner detector with $p_T>6$~GeV and $|\eta|<2.5$. (b) The total electric charge in each hemisphere (as defined by the trigger muon direction).}
\label{fig:nc46}
\end{figure}

The second line of the table illustrates the effect of demanding only 4 or 6 charged tracks. Specifically, for each charged particle, we evaluate the probability that the particle is reconstructed as a track by the ATLAS inner detector (ID) and remove particles from the event based on this probability.  We also only consider those particles with $p_T > 500$~MeV and pseudo-rapidity satisfying $|\eta| < 2.5$, which is the approximate acceptance of the ATLAS ID. All track kinematics are smeared with detector resolution and each track assigned a vertex using the ID vertex resolution. Each track vertex is then required to lie within 2.5~mm of the vertex defined by the muon. This last requirement is important in minimizing the impact of charged tracks from pile-up events. The number of charged particles within the vertex window, for zero pile-up events, is shown in Figure~\ref{fig:nc46}(a).

A smaller vertex window is not desirable because as the vertex window
decreases there is an increasing possibility that, after vertex
smearing, charged particles are assigned a vertex outside of the
window and would be lost to the analysis thereby reducing the
signal. Further, because the background events typically have greater than 6 charged
particles in the primary event, the background can even increase as this window decreases. 
Of course, taking too large a window allows pile-up to contaminate the signal\footnote{It contaminates the background too.}. We choose 2.5~mm as a compromise and show the effect of changing the window size in Section \ref{pile}.

The event is then divided into hemispheres, defined by the axis of the highest-$p_T$ muon. Figure~\ref{fig:nc46}(b) shows the sum of the electric charges in each hemisphere for the signal and background events. We assume 100\% efficiency in assigning the charge of each track, which is a reasonable assumption for reconstructed low transverse momentum tracks as the curvature of the track due to the magnetic field will be well measured. The third line of Table \ref{tb:sigmas} follows from the requirement that there is no net charge in each hemisphere.

\begin{figure}[t]
\centering
\mbox{
        \subfigure[]{\includegraphics[width=0.5\textwidth]{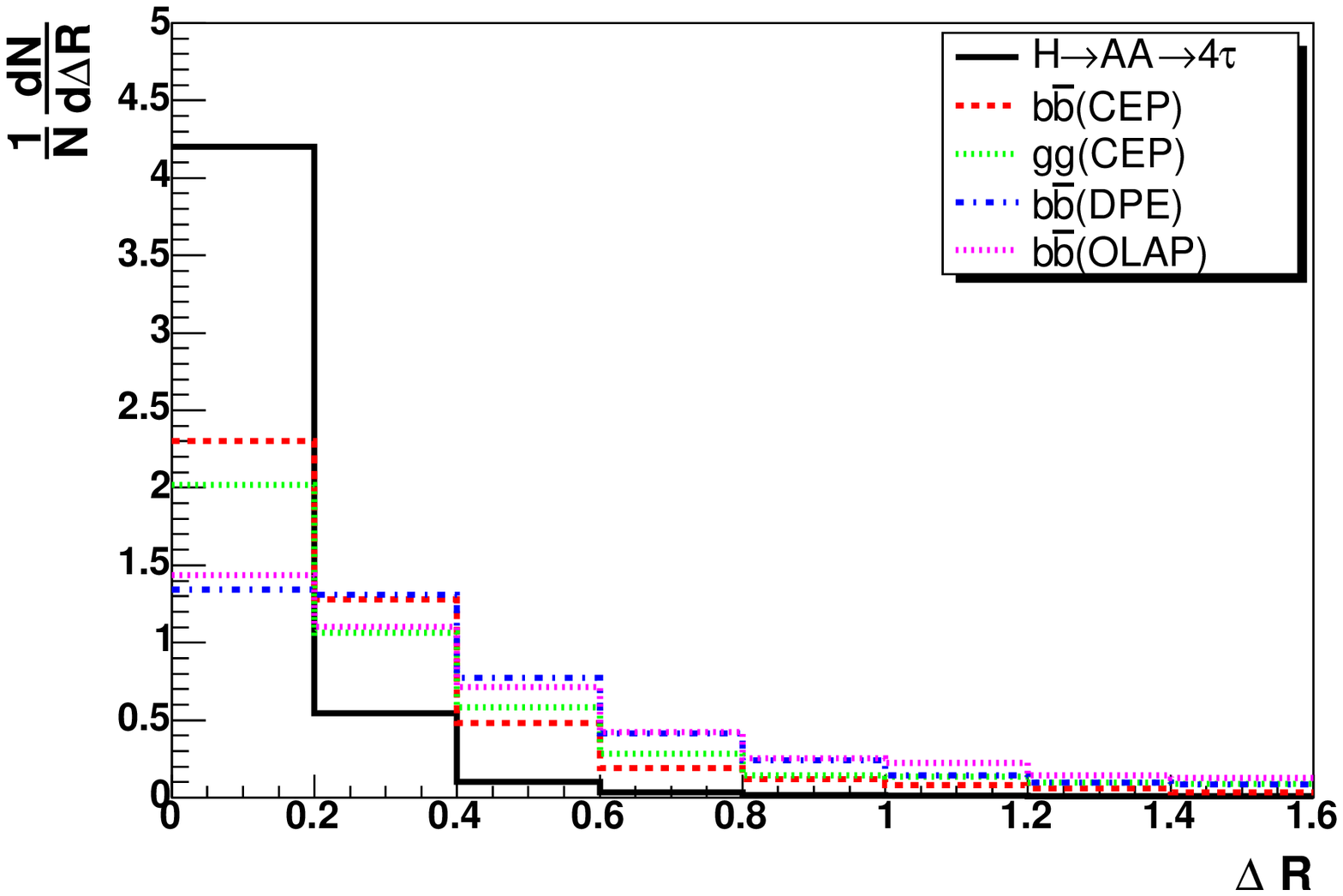}}
        \subfigure[]{\includegraphics[width=0.5\textwidth]{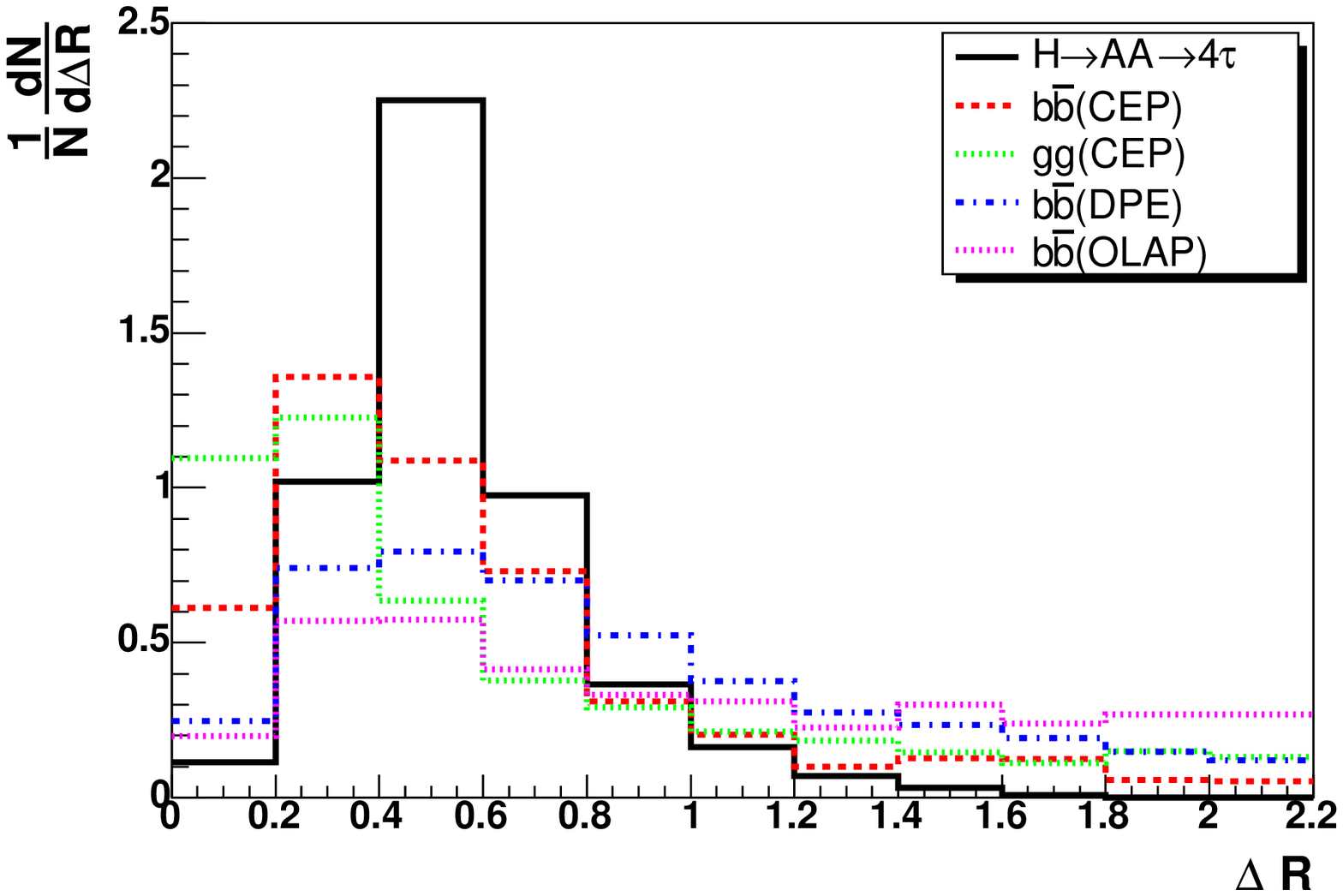}}
}
\caption{(a) The distance in $\eta\times\phi$ of the three clustered tracks (in a 6 track event) that define a single tau object from the weighted average of the tracks. (b) The distance in $\eta\times\phi$ of each tau object from its nearest neighbor.}
\label{fig:deltaR}
\end{figure}

The next cut, the topology cut, requires a little explanation. The
first step is to define four `tau objects'. In 4-track events this is
trivial but for 6-track events we look to merge 3 tracks into a single
object. The algorithm works as follows:
\begin{enumerate}
\item The nearest two tracks are found using the measure
\begin{equation}
r_{ij}^2 = (\eta_i - \eta_j)^2 + (\phi_i - \phi_j)^2
\end{equation}
and merged into a single object with a pseudo-rapidity and azimuth given by the $p_T$ weighted average of the pair,
{\it i.e.} 
\begin{equation}\label{eq:wgtav}
\eta = \frac{\displaystyle\sum_{i} p_{Ti}  \, \eta_i}{\displaystyle\sum_{i} p_{Ti}}
\end{equation}
and similarly for $\phi$.  
\item The nearest track to this pair is found and a tau object defined by clustering the 3 tracks together. The direction of the corresponding tau object is given by the $p_T$ weighted average of the pseudo-rapidity and azimuth of the 3 tracks given by (\ref{eq:wgtav}). The $p_T$ of the object is just the sum of the individual track transverse momenta.
\end{enumerate}
Figure~\ref{fig:deltaR}(a) shows the distance in $\eta\times\phi$ space of the three tracks from the newly defined tau object and events are only retained if each of the three tracks lies within $\Delta R<0.2$ of the tau object direction. For the surviving events we pair up the four tau objects. Figure~\ref{fig:deltaR}(b) shows the distance in $\eta\times\phi$ space of each tau object from its nearest neighbor.  We exploit the fact that the signal has two pairs of roughly collinear
tracks by cutting on the separation between the tracks which form each pair, {\it i.e.} $\Delta R < 1$.

\begin{figure}[t]
\centering
\mbox{
        \subfigure[]{\includegraphics[width=0.5\textwidth]{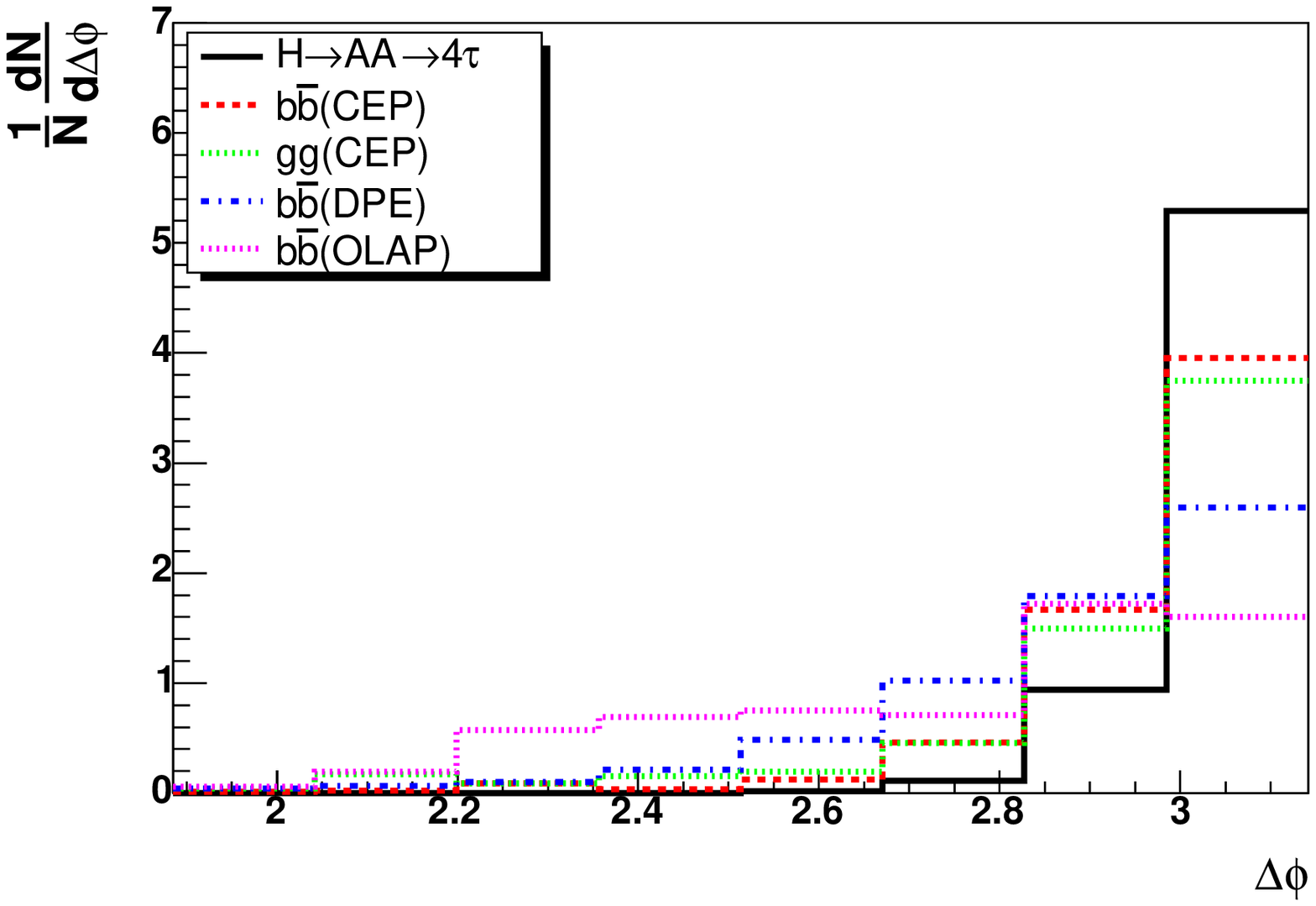}}
        \subfigure[]{\includegraphics[width=0.5\textwidth]{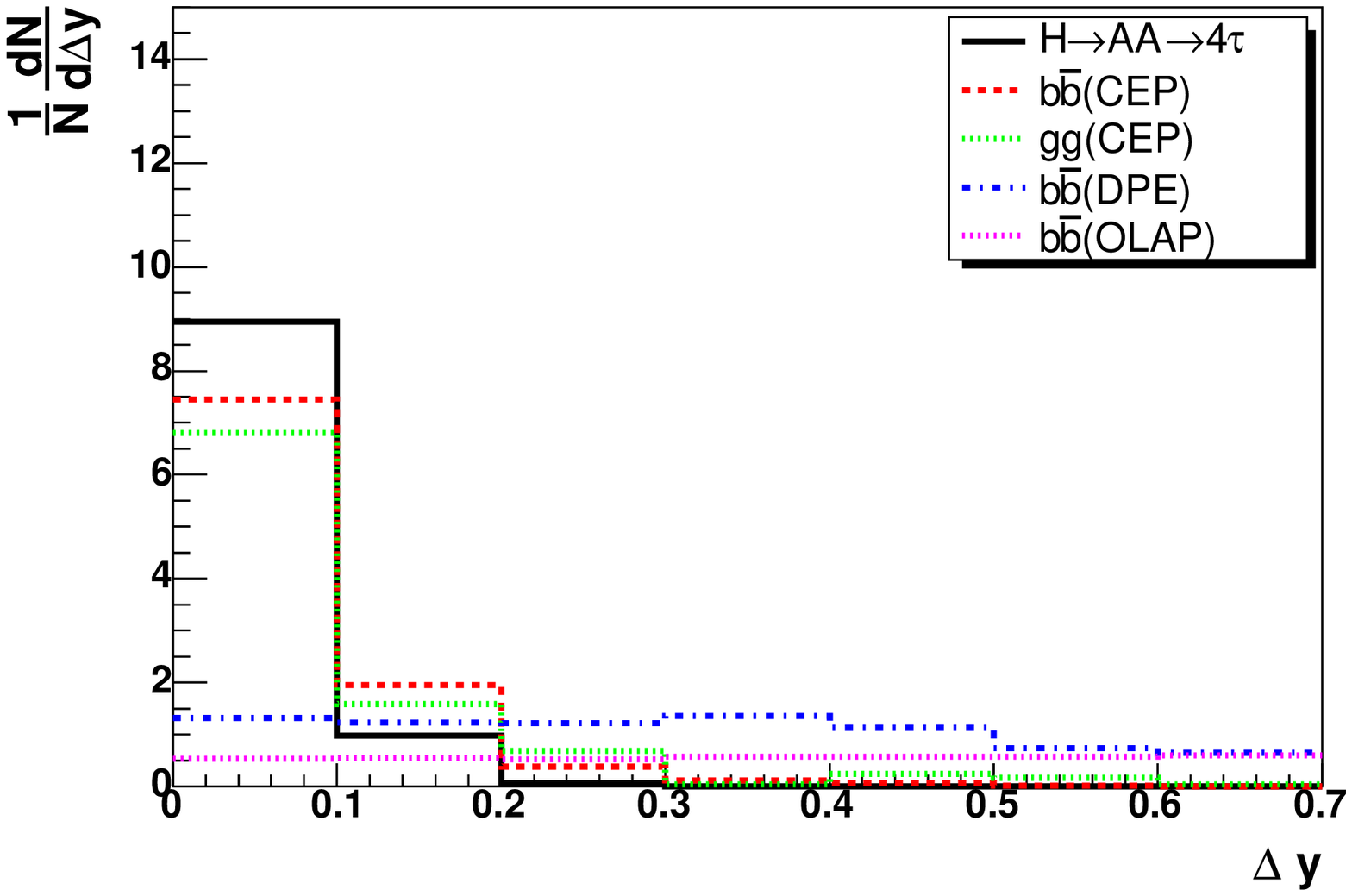}}
}
\caption{(a) The difference in azimuth between the reconstructed pseudo-scalar objects. (b) The $\Delta y$ distribution.}
\label{fig:pseudoscalarcuts}
\end{figure}

The pairing up of the tau objects corresponds to identifying the collinear decay products of each of the two pseudo-scalars. We refer to each pair as a pseudo-scalar object which is constructed using the $p_T$ weighted average $(\eta,\phi)$ for each pair of tau objects. The $p_T$ of each pseudo-scalar object is just the sum of the $p_T$'s of the constituent tau objects. Figure~\ref{fig:pseudoscalarcuts}(a) shows the azimuthal separation of each pseudo-scalar object, for the signal and background events, and we make a cut on $\Delta \phi > 2.8$ to enforce that they are back-to-back. Finally, we
make the cut
\begin{equation} \label{eq:dely}
\Delta y = \frac{1}{2} \left| \log(\xi_1/\xi_2) - (\eta_1 + \eta_2) \right| < 0.1
\end{equation}
where $\xi_i$ is the fractional momentum lost by incoming proton $i$
(as determined by the 420~m detectors). This cut exploits the fact
that the forward detectors are able to determine the boost which will put the
central system into the zero momentum frame and that in this frame the
primary $a$ particles (and hence their approximately collinear decay
products) have rapidities that are almost equal in magnitude (but
opposite in sign). Figure~\ref{fig:pseudoscalarcuts}(b) shows the $\Delta y$ variable for the signal and background events. The net result of these topology cuts is presented in the 4th line in the table.

\begin{figure}[t]
\centering
\mbox{
        \subfigure[]{\includegraphics[width=0.5\textwidth]{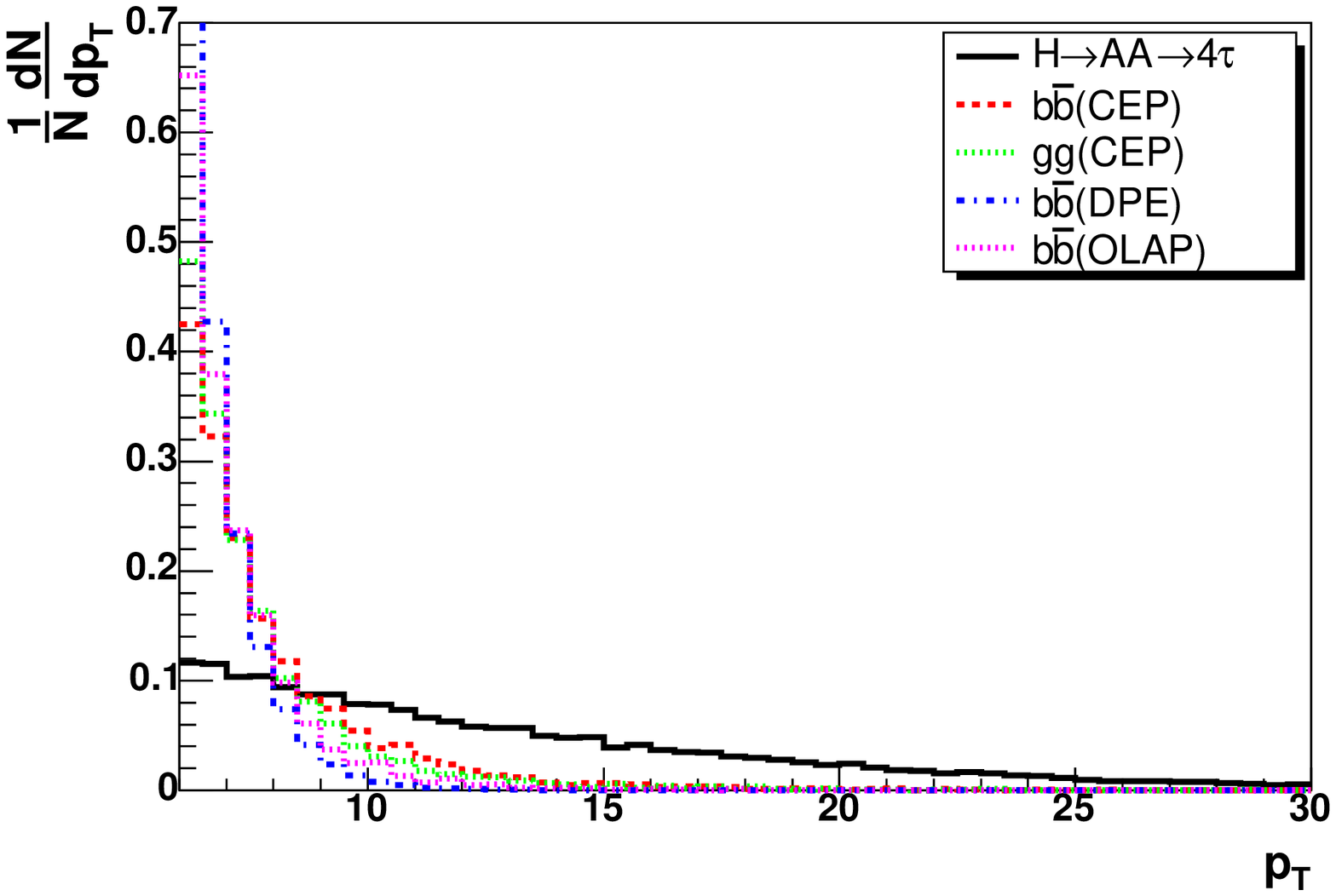}}
        \subfigure[]{\includegraphics[width=0.5\textwidth]{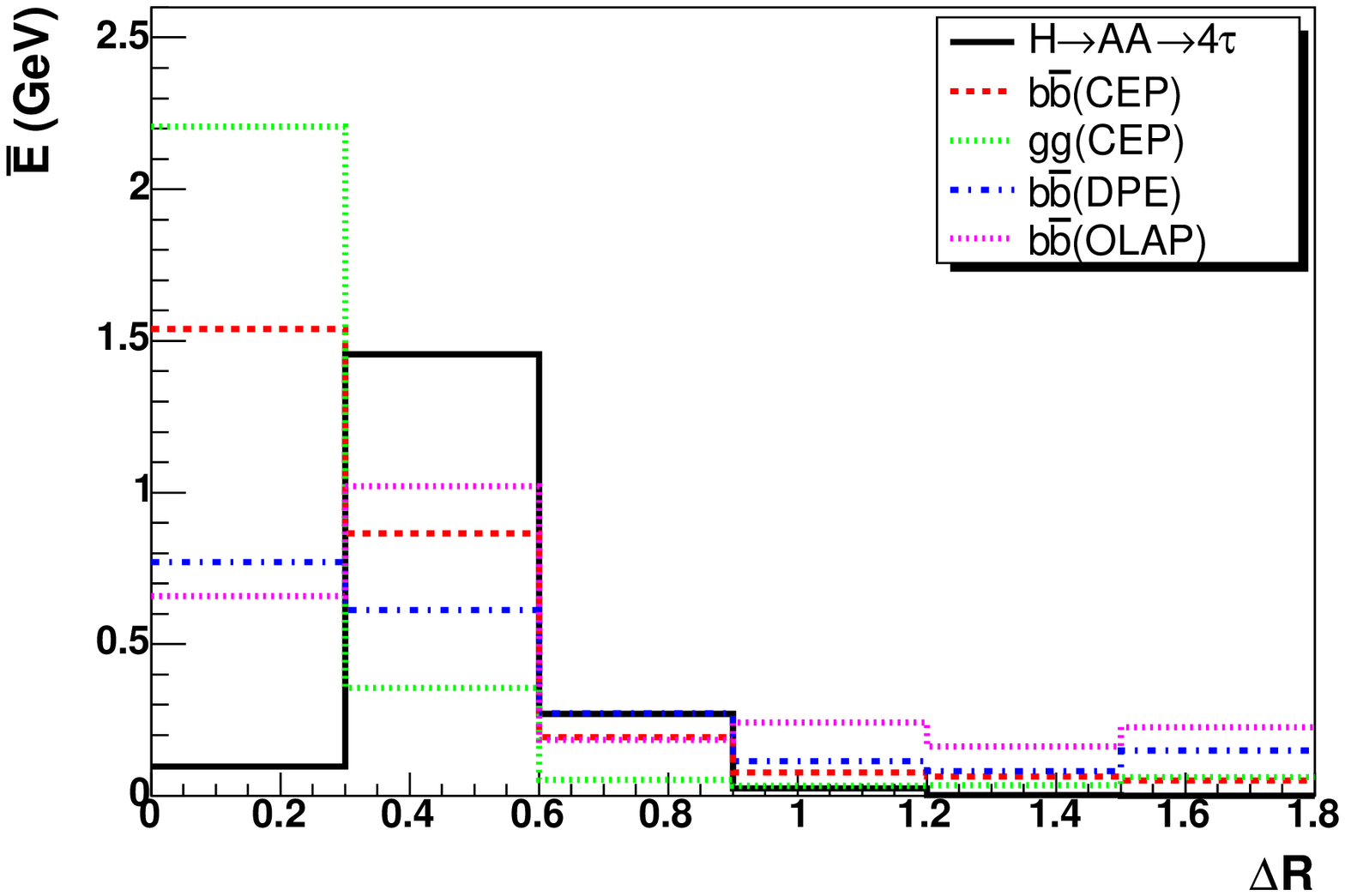}}
}
\caption{(a) The $p_T$ of the trigger muon. (b) The average transverse energy deposited in the calorimeter within a distance, $\Delta R$, in $\eta \times \phi$ space of the trigger muon.}
\label{fig:muisol}
\end{figure}

We have already insisted upon a single muon with $p_T > 6$~GeV. This value is chosen since it is the smallest level 1 muon trigger threshold foreseen at ATLAS. Figure~\ref{fig:muisol}(a) shows the $p_T$ spectrum of the trigger muon after the charged track multiplicity cut: it suggests we should use a higher cut of $p_T > 10$~GeV. In addition to this, we require that the muon be isolated, {\it i.e.} that there is no hadronic activity near to the muon.
Figure~\ref{fig:muisol}(b) shows the average transverse energy deposited within a distance $\Delta R$ of the muon, after the muon $p_T > 10$ GeV cut. We require that the transverse energy in the calorimeter lying within $\Delta R < 0.3$ of the muon be less than 2 GeV (excluding the $\mu$). This allows for the elimination of backgrounds in which the muon is produced as a result of in-jet particle decays, {\it i.e.} in particular the $gg$ backgrounds. We examine the isolation criterion further in Section \ref{pile} since it is sensitive to pile-up. Note that this is the only cut we use which makes use of the calorimeter.

\begin{figure}[t]
\centering
        \includegraphics[width=0.5\textwidth]{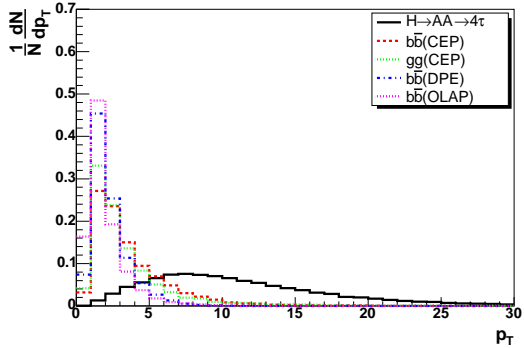}
\caption{ The $p_T$ of the highest momentum track that is not the trigger muon}
\label{fig:pt2}
\end{figure}

We next cut on $p_T^{1, \not{\mu}}$, which is the transverse momentum of the highest $p_T$ track that is not the trigger muon. Figure~\ref{fig:pt2} shows the value of $p_T^{1, \not{\mu}}$ for the signal and background events before the topology cuts. We choose to present this distribution at this point because the statistics are large for each sample. However, in the table the cut is applied after the topology cut. The background particles have mainly low $p_T$ tracks originating from the hadronization of jets. The signal tracks however, originate from tau decays and can have much larger momentum. We make the cut  $p_T^{1, \not{\mu}}>6$~GeV in line 6 of the table. 

\subsection{Reconstruction of the Higgs masses}\label{sec:masses}

Since the $a$ bosons are much lighter than the $h$, it follows that they are highly boosted and that their decay products will travel along the direction of the originating $a$. This means that the four-momentum of the primary $a$ is, to a fair approximation, directly proportional to the visible four-momentum of the charged particles into which it decays, {\it i.e.}
\begin{equation}
p_{i}^{vis} = f_i \; p_{a,i}
\end{equation}
where $p_i^{vis}$ is the visible four-momentum. Since we are still
restricting our analysis to charged tracks, $p_i^{vis}$ is missing the
neutrino momentum and the momentum of any neutral particles. $p_{a,i}$
is the four-momentum of the corresponding pseudoscalar $a$ and $1-f_i$
is the fraction of the $a$ momentum carried away by neutrals.  We
therefore have only two unknowns, $f_1$ and $f_2$ and these can be
determined using information from the forward proton detectors since
we know that
\begin{equation}
p_{a,1} + p_{a,2} = p_h
\end{equation}
and $p_h$ can be measured. In fact the situation is over-constrained since we have only two unknowns and three equations. Although the transverse momentum of the Higgs can in principle be measured using the forward detectors it will typically be rather small. Assuming it to be zero leaves us with the three equations: 
 
\begin{equation}\label{eq:recon1}
\frac{(p^{vis}_1)_{x,y}}{f_1} + \frac{(p^{vis}_2)_{x,y}}{f_2} = 0
\end{equation}   
and
\begin{equation} \label{eq:recon2}
 \frac{(p^{vis}_1)_z}{f_1} + \frac{(p^{vis}_2)_z}{f_2}=(\xi_1 - \xi_2) \frac{\sqrt{s}}{2} 
\end{equation}   
where $x$ and $y$ label the directions transverse to the beam axis and the $1-\xi_i$ are the longitudinal momenta of the outgoing protons expressed as fractions of the incoming momenta. Solving~(\ref{eq:recon1}) and~(\ref{eq:recon2})  gives 
\begin{eqnarray} \label{eq:recon3}
f_1 &=& \frac{2}{(\xi_1 - \xi_2) \sqrt{s}} \left[ (p^{vis}_1)_z - \frac{(p^{vis}_2)_z (p^{vis}_1)_{x,y}}{(p^{vis}_2)_{x,y}} \right]\; ,
\\ 
\label{eq:recon4}
f_2 &=& - \frac{(p^{vis}_2)_{x,y}}{(p^{vis}_1)_{x,y}} f_1 \; .
\end{eqnarray}   
Equations (\ref{eq:recon3}) and (\ref{eq:recon4}) provide two solutions depending on whether we solved using the  $(x,z)$ or $(y,z)$  pair of equations. We therefore have 4 determinations of the $a$ mass per event.\footnote{The $(x,y)$ pair of equations could also be used if one could make a good measurement of the outgoing proton transverse momenta.}

\begin{figure}[t]
\centering
\mbox{
        \subfigure[]{\includegraphics[width=0.5\textwidth]{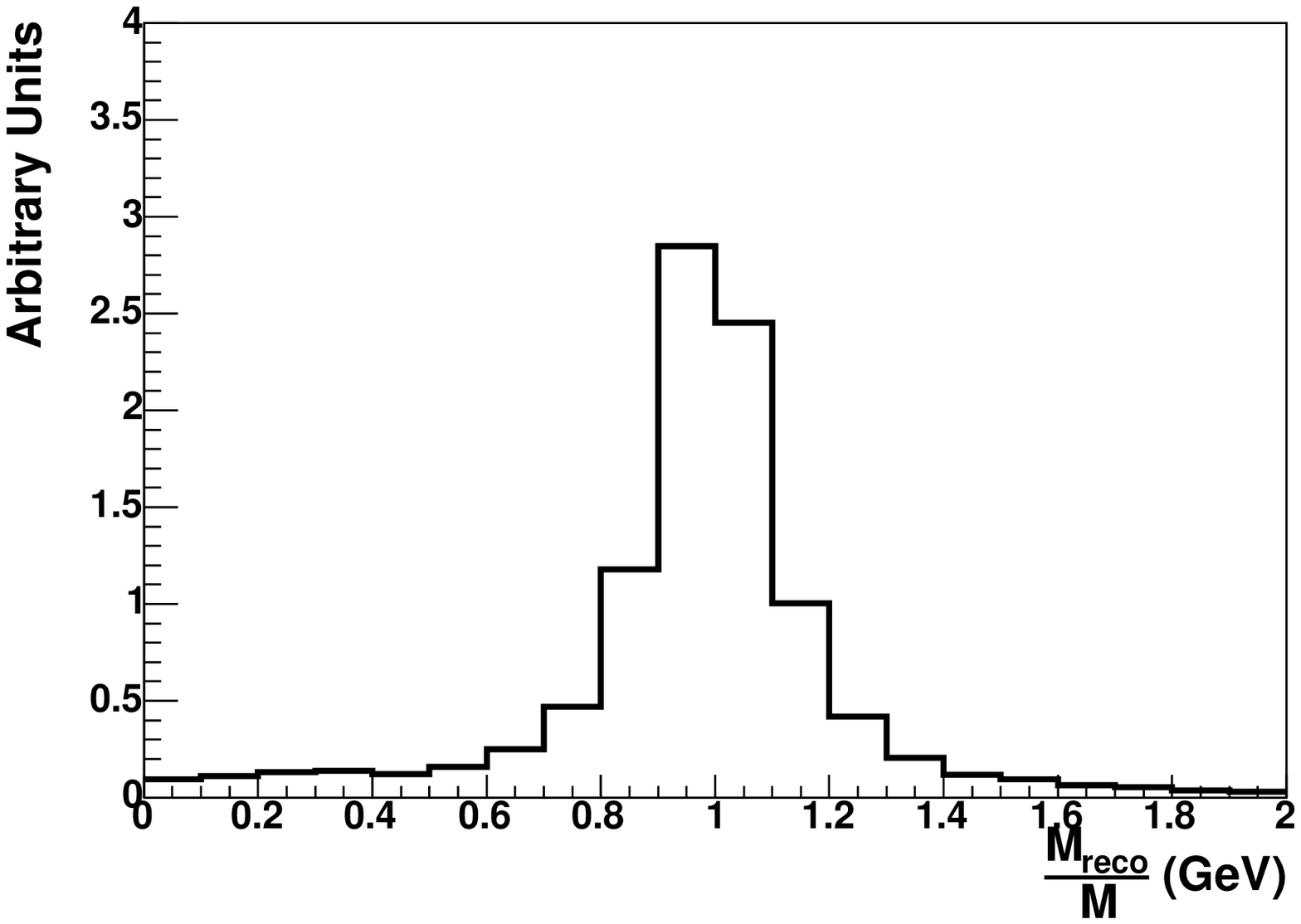}}
        \subfigure[]{\includegraphics[width=0.5\textwidth]{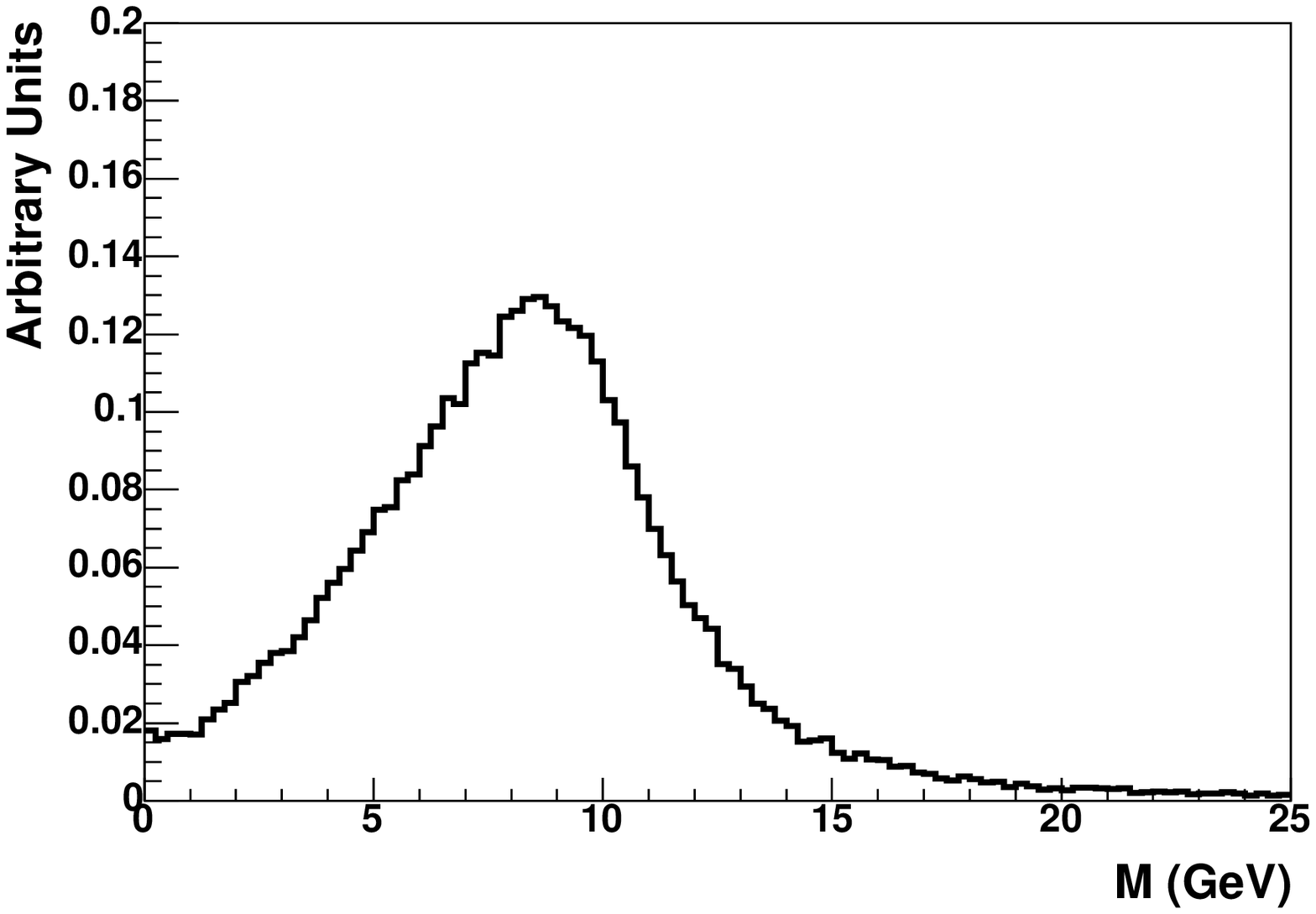}}
}       
\caption{(a) The ratio of the reconstructed scalar Higgs mass to the mass measured by FP420 for the signal events only. (b) The reconstructed $a$ mass for the signal events. The broad distribution is due to the breakdown of the collinearity approximation and detector effects are minimal.}
\label{fig:massH}
\end{figure}

Using this method, we are able to reconstruct the masses of both the
scalar Higgs and the pseudo-scalar Higgs. Figure~\ref{fig:massH}(a)
shows the ratio of the reconstructed scalar Higgs boson mass to the
mass measured in the forward detectors for the signal. The distribution is broad,
mainly as a result of the collinearity approximation and the missing
momentum carried by neutral particles (this is a bigger effect than
that due to detector smearing). In the final analysis, we do not need
to apply a cut on this variable because we have already adequately
reduced the background. In Figure~\ref{fig:massH}(b), we show the
pseudo-scalar mass distribution using only signal events. 
The final cut in Table \ref{tb:sigmas} corresponds to restricting the 
mean of each of the two reconstructed $a$ masses to the range $m_a > 2 m_\tau$.

\begin{figure}[t]
\centering
\includegraphics[width=0.5\textwidth]{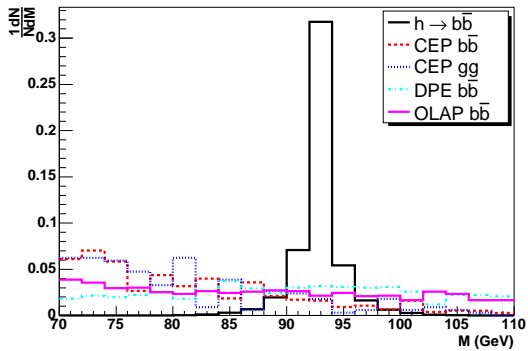}
\caption{The mass distribution of events after the topology requirement for signal and selected background.}
\label{fig:mass}
\end{figure}

Finally, it should be noted that we have not yet applied a stringent mass window requirement. The background was generated with a central mass in the range \linebreak 70 GeV~ $<M<110$~GeV and the mass distribution as measured by FP420 (with detector smearing) is shown in Figure~\ref{fig:mass} after the application of the topology requirement. It is clear that in estimating the significance we should further limit ourselves to a smaller region of interest around the signal.

\subsection{Estimating non-$b\bar{b}$ backgrounds} \label{sec:rescale}
In Table \ref{tb:sigmas} the OLAP background quoted is obtained by superimposing an inclusive $pp \to b \bar{b}+X$ event with two single diffractive events $pp \to p+X$. As stated, we obtain the full OLAP background after re-scaling the OLAP background from this source by a factor  $\sigma(pp \to jj+X)/\sigma(pp \to b\bar{b}+X)$. PYTHIA is used to generate the events and the cross-sections are obtained after insisting that there be at least one muon with $p_T>6$ GeV, and after the proton acceptance and the charged track cuts: we find that a re-scaling of the $b\bar{b}$ OLAP background by a factor 5.0 is necessary.

The re-scaling is determined after the charged track cut since many of the muons produced in the $jj+X$ sample arise from processes involving gluon jets (this is certainly so in the case of  $gb \to gb$ intrinsic $b$ contributions) in which case there are typically more tracks than in a typical $b\bar{b}$ event and hence the track cut should be more effective at reducing this source of background. The results quoted in the following sections include this re-scaling.

A similar re-scaling is performed, this time using POMWIG, in order to
convert the DPE $b\bar{b}$ background to the full $jj$ background and
we find that a correction factor of 1.86 is required in this case. This number is smaller than
for the inclusive case since the pomeron is gluon dominated: In the inclusive case, intrinsic $c$ and $b$ quark contributions actually dominate the production of those dijet events which pass our muon $p_T$ and charged track cuts, via the process $gq \to gq$.

One may also worry about OLAP coming from processes other than QCD dijet production. For example, the inclusive QED production of four taus in conjunction with two single diffractive events. However, this background is negligible since we know that the exclusive QED background is negligible, as demonstrated in Table 1 (and this was computed without any form factor suppression). The fact that the photons can couple to quarks within the protons is not able to compensate for the suppression factor of over two orders of magnitude arising from the requirement that there also be two single diffractive events which produce a fake vertex coincident with the vertex from the muon.

\subsection{Triggering}\label{trigger}

In this section, we discuss the trigger options and trigger
efficiencies for this analysis. The signal from FP420 will not arrive
at the central trigger processor in time for the level 1 decision. At
level 2 and above however, we can make use of the FP420 and the muon vertex information to
substantially reduce the rate; requiring two in-time proton hits
results in a reduction factor of 140 at high luminosity and
approximately 20000 at low luminosity \cite{Cox:2007sw}. We therefore
focus on the possible ways to retain the events at level 1.

As we already stated, our strategy is to trigger on a muon arising from the decay of one of the taus. Before any analysis cuts, we imposed a minimum muon $p_T$ requirement of 6~GeV. This is the lowest trigger threshold foreseen at ATLAS and it would be pre-scaled at high luminosity in order to reduce the rate and may even be pre-scaled at low luminosity. However, in our analysis we make a further cut on the muon; requiring $p_T>10$~GeV in order to improve the significance of any observation. In our subsequent analysis, we will therefore employ what we call the MU10 trigger, {\it i.e.} a muon trigger threshold of 10~GeV. For muons with $p_T > 10$ GeV, the trigger efficiency is 80\%  \cite{triggerconf1,triggerconf2}; in this way we avoid the need to pre-scale without losing too much signal.
It should be noted that we still need to employ the $p_T > 10$ GeV analysis cut since the MU10 trigger does not directly measure  the muon $p_T$, which results in the acceptance of some muons with $p_T$ below 10 GeV. The bottom line is that, for the MU10 trigger, we multiply the event rates in Table \ref{tb:sigmas} by 0.8 to reflect the ATLAS trigger efficiency.

The fact that the ATLAS level 1 muon trigger has acceptance below the nominal threshold allows us a second possibility to reduce rates: we could use a MU15 trigger. Such a trigger would have an efficiency of 80\% for muons with $p_T > 15$~GeV, but would retain some efficiency for $p_T$ values in the region $10-15$~GeV. In \cite{triggerconf2}, results are presented for the L1 muon trigger efficiencies for 11 GeV and 20 GeV thresholds and from these results it is apparent that the efficiency for triggering on 10~GeV muons given a 15~GeV threshold will be between 20\% and 50\%. We choose to model the efficiency of a MU15 trigger with a linear turn-on, with 20\% efficiency at 10~GeV and 80\% efficiency at 15~GeV. Above 15~GeV we assume the canonical 80\% efficiency. This linear turn-on will probably underestimate the true efficiency because the real trigger should be more efficient in the region near to the threshold (and because the trigger could in fact be closer to 50\% efficient for 10~GeV muons). The cut on muon $p_T$ at analysis level can therefore remain at 10~GeV even with a MU15 trigger.
We note that any selection bias due to a different $\eta-\phi$ dependence of the signal and
background is expected to be small since they have rather similar behaviour in these variables.
In any case, this small effect would not affect our estimate of the significance provided one can
determine the background by extrapolating from outside the signal region.

\subsection{Contamination from pile-up}\label{pile}

\begin{figure}[t]
\centering
\mbox{
        \subfigure[]{\includegraphics[width=0.5\textwidth]{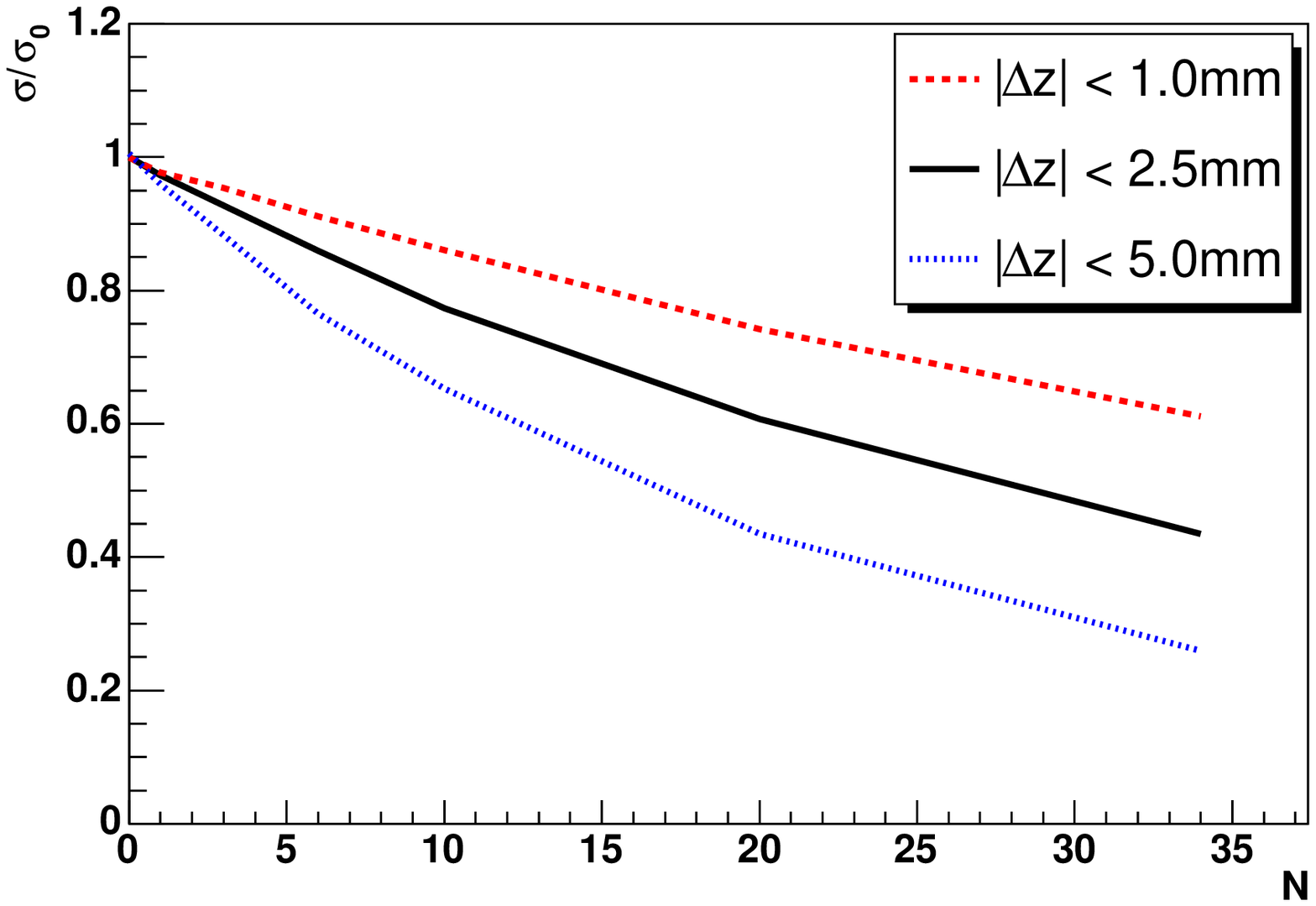}}
        \subfigure[]{\includegraphics[width=0.5\textwidth]{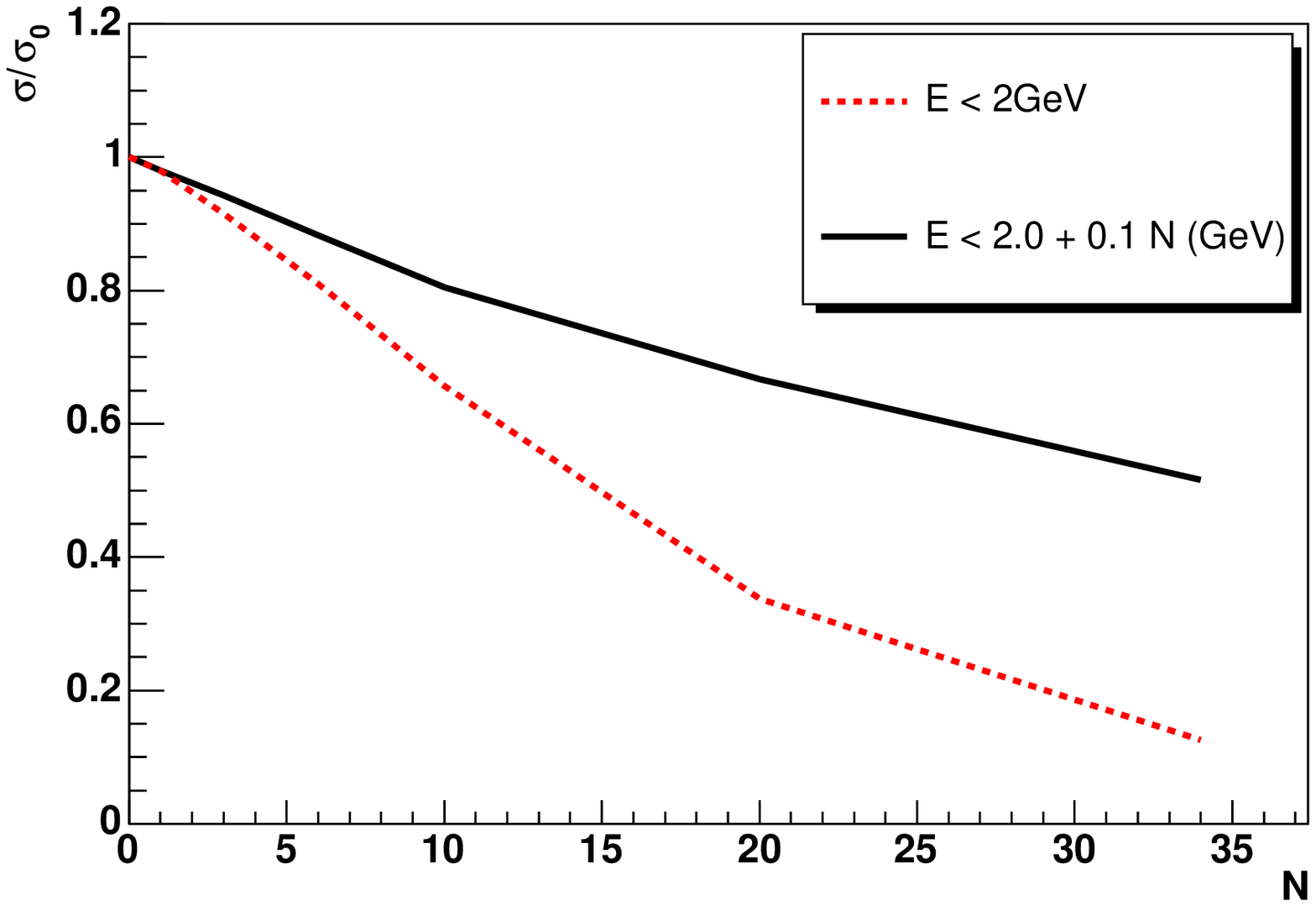}}
}       
\caption{(a) The fractional reduction in the signal cross-section
  (evaluated after the $p_T^{1, \not{\mu}}$ cut)  as a function of the
  number of pile-up events for three different vertex window
  sizes. (b) The fractional reduction in the signal cross-section as a function of the number of pile-up events for two different muon isolation criteria.}
\label{fig:pileeffect}
\end{figure}

So far we have assumed that the contamination of both signal and
background events from pile-up is negligible. In this section, we
quantify the effect of pile-up on our estimates by adding events
generated by PYTHIA to our signal events. The first effect is that
charged tracks from the pile-up events do cause signal events to fail
the charged track cut when the pile-up vertex is near to the primary
vertex. Figure~\ref{fig:pileeffect}(a) shows the fractional reduction
in signal for various $\Delta z$ choices due to the extra charged
tracks from pile-up as a function of the number of pile-up events
overlaid with the signal event. The cross-section is relative to that
obtained without pile-up for the given $\Delta z$ and is evaluated
after the $p_T^{1, \not{\mu}}$ cut. The effect of changing the vertex
window size is also shown. Note that, as discussed in Section
\ref{analysis}, we choose to use $|\Delta z| < 2.5$ mm in order to
avoid the  increase in background, which arises as $|\Delta
z|$ decreases, and also to avoid having too many signal tracks start
to fall outside the window.

The other effect of pile-up is that the additional energy in the calorimeters can cause events to fail the muon isolation cut. Figure~\ref{fig:pileeffect}(b) shows the reduction in signal events due to the increased energy in the calorimeter as a function of the number of pile-up events. It is clear that requiring $E_T<2$~GeV in the calorimeter within $\Delta R < 0.3$ of the muon is only valid at low luminosities. Figure~\ref{fig:pileeffect}(b) also shows the effect of imposing a luminosity dependent isolation cut given by
\begin{equation}
E_T <  2.0 + \frac{\bar{N}}{10}   \quad \rm{(GeV)} \label{eq:muiso}
\end{equation}
where $\bar{N}$ is the average number of pile-up events in a bunch
crossing. This results in an isolation requirement of 5.4 GeV at high
luminosity, which retains approximately 50\% of the signal. In what follows we use Eq.~(\ref{eq:muiso}).

\subsection{The significance}\label{signif}

In this section we make predictions for the significance of any observation given three years of data acquisition at a fixed luminosity.  We take the nominal integrated data acquisition rate of 10~fb$^{-1}$~yr$^{-1}$ at low luminosity ($10^{33}$~cm$^{-2}$~s$^{-1}$) and 100~fb$^{-1}$~yr$^{-1}$ at high luminosity ($10^{34}$~cm$^{-2}$~s$^{-1}$).  In order to predict final event rates, we must re-scale the results presented in Table~\ref{tb:sigmas} as follows:
\begin{itemize}
\item The DPE and OLAP $b\bar{b}$ backgrounds are scaled up as explained in Section \ref{sec:rescale};
\item All event rates are scaled to reflect ATLAS trigger efficiency as explained in Section~\ref{trigger};
\item All event rates are scaled to reflect the effect of pile-up on the charged track multiplicity and the muon isolation cuts, {\it i.e.} using the solid black curves in Figures \ref{fig:pileeffect}(a) and (b);
\item For significance purposes, we are interested in the event rates in a region around the Higgs mass peak of $m_h\pm5$ GeV. We reduce our background rates accordingly.
\end{itemize}

Table \ref{eventnumbers} shows the final number of signal and background events after three years of data acquisition at an instantaneous luminosity of $10^{33}$, $5\times10^{33}$ and $10^{34}$~cm$^{-2}$~s$^{-1}$. We present the final event numbers for the two trigger strategies, MU10 and MU15, and also for MU10 with reduced overlap backgrounds, which assumes that 2~ps proton time-of-flight can be achieved by FP420. 

\begin{table}[t]
\begin{center}
\begin{tabular}{|c|c|c||c|c||c|c|}
\hline

 Luminosity & \multicolumn{2}{c|}{MU10} &  \multicolumn{2}{c|}{MU15} &  \multicolumn{2}{c|}{MU10 (2~ps)} \\
  ($\times10^{33}$~cm$^{-2}$~s$^{-1}$)  & S & B & S & B & S & B \\
 \hline
 $1$& 1.3 & 0.02 & 1.0 & 0.01 & 1.3 & 0.02\\
 $5$ & 3.7 & 0.14 & 2.9 & 0.08 & 3.7 & 0.07\\
 $10$ & 3.3 & 0.36 & 2.5 & 0.20 & 3.3 & 0.11\\
\hline
\end{tabular}
\end{center}
\caption{Expected number of signal (S) and background (B) events for the three trigger scenarios assuming that the data are collected at a fixed instantaneous luminosity over  a three year period. We assume the integrated luminosity acquired each year is 10~fb$^{-1}$, 50~fb$^{-1}$ and 100~fb$^{-1}$ at an instantaneous luminosity of 1$\times10^{33}$~cm$^{-2}$~s$^{-1}$, 5$\times10^{33}$~cm$^{-2}$~s$^{-1}$ and 10 $\times10^{33}$~cm$^{-2}$~s$^{-1}$. \label{eventnumbers}}
\end{table}

 We estimate the significance, $S$, of the observation using
\begin{equation}
\frac{1}{\sqrt{2\pi}} \int_{S}^{\infty} \mathrm{e}^{\frac{-x^2}{2}} dx  =  \sum_{n=s+b}^{\infty} \frac{b^n \mathrm{e}^{-b}}{n!} 
\end{equation}
where $s$ is the number of signal events and $b$ is the number of background events. The probability of observing $s+b$ events given a Poisson distribution with mean $b$ is equated to the probability of  $S$ standard deviations in a Gaussian distribution. 

\begin{figure}[t]
\centering
\mbox{
        \subfigure[]{\includegraphics[width=0.5\textwidth]{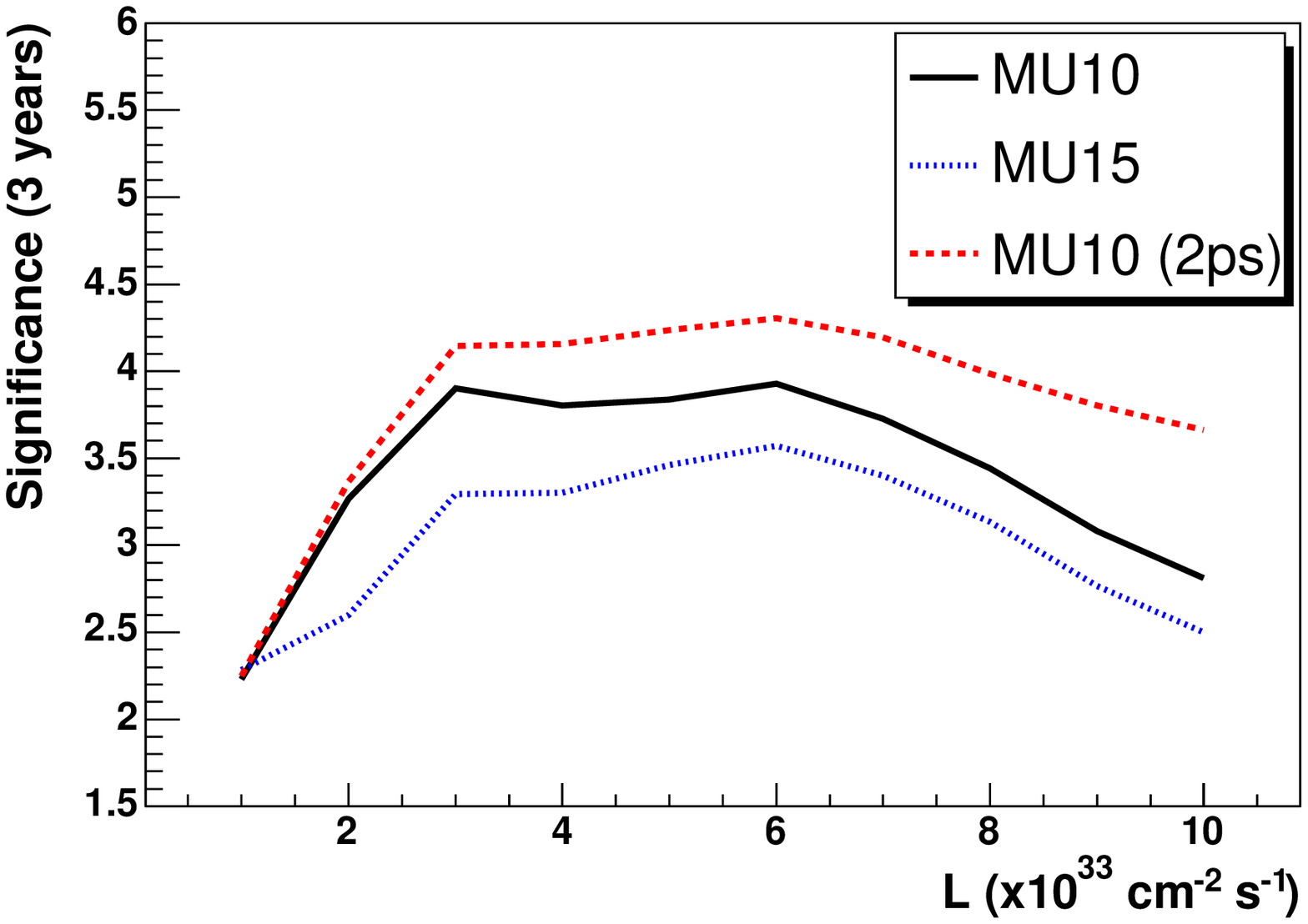}}
        \subfigure[]{\includegraphics[width=0.5\textwidth]{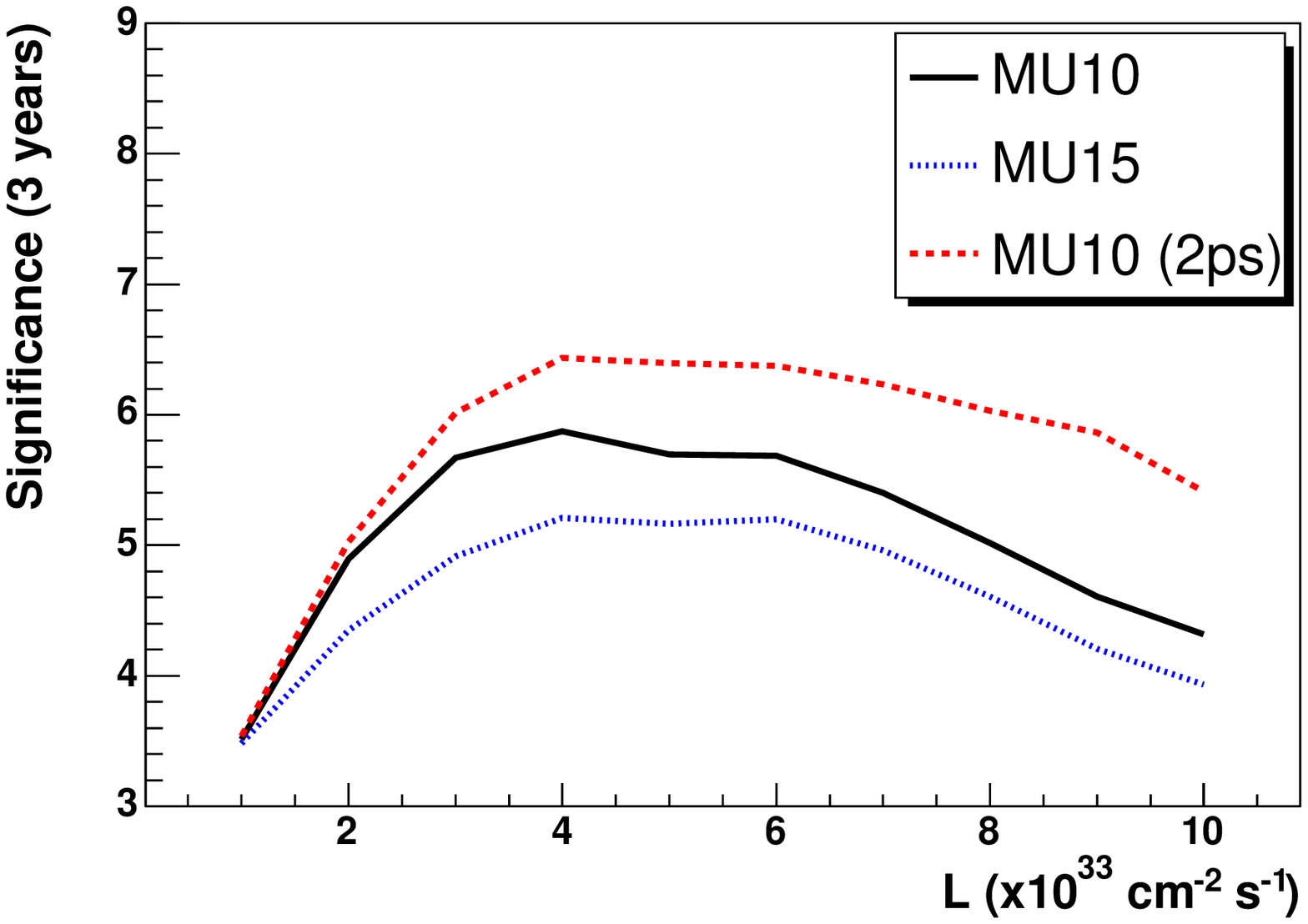}}
}
\caption{(a) The significance for three years of data acquisition at each luminosity. (b) Same as (a) but with twice the data.}
\label{fig:signif}
\end{figure}

Figure~\ref{fig:signif}(a) shows the significance of the measurement after three years of data acquisition at each luminosity. The lower curves correspond to the two different muon $p_T$ trigger thresholds whilst the upper curve shows what it possible if the forward detector timing accuracy can be improved from 10~ps to 2~ps.
Figure~ \ref{fig:signif}(b) is the same as Figure~\ref{fig:signif}(a) but assuming that twice the amount of data are available, {\it e.g.} as might occur if the results of ATLAS were combined with those of CMS. 

\begin{figure}[t]
\centering
\includegraphics[width=0.5\textwidth]{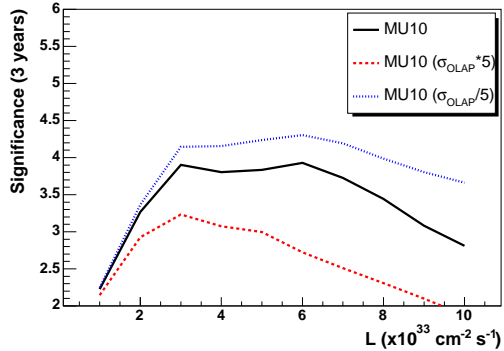}
\caption{The significance of the measurement for reduced overlap (OLAP) contributions.}
\label{fig:signifnolap}
\end{figure}

Figure~\ref{fig:signifnolap} shows the significance of the measurement
if the overlap background was increased or decreased by a factor of 5,
assuming the nominal data acquisition rate and 10~ps time-of-flight
measurements. This plot illustrates the importance of the OLAP
background at higher luminosities. We observe that a five-fold
increase in the OLAP background would result in the analysis being
possible only at low luminosity. A similar plot for the case of twice
the data being available shows that the OLAP background would have
to be more than a factor of 40 larger than our best estimate before
the signal would become unobservable. We show Figure
\ref{fig:signifnolap} because we do not believe it is possible to make
a precise prediction for the OLAP background given current
understanding. In particular we note that the huge reduction factors
for this background arise mainly because we make cuts on the charged
track multiplicity, and this is heavily dependent on the underlying
event. In this analysis, we chose to use the `ATLAS tune' of PYTHIA
which fits the Tevatron data. However, the extrapolation of these fits
to LHC energies is uncertain and different models/tunes predict a
different amount of underlying event activity. For example, if we had
chosen to use the HERWIG event generator  \cite{Corcella:2002jc}, with JIMMY \cite{Butterworth:1996zw}  simulating the
underlying event, the charged track activity would be
increased  \cite{jimmytune} which would make the overlap events more likely to fail the
charged track multiplicity cuts thereby reducing the OLAP background.
Apart from uncertainties arising due to our lack of understanding of
the OLAP background, our estimates have assumed that all of the data
are collected at a value of the instantaneous luminosity fixed to the
canonical value ($10^{33}$ cm$^{-2}$s$^{-1}$ or $10^{34}$cm$^{-2}$s$^{-1}$). In reality, the luminosity decreases from the
canonical value with time until the next LHC fill. The decay time of
the luminosity at the LHC is expected to be 14.9 hours \cite{Bruning:2004ej} and the ratio
of the average luminosity per fill to the initial luminosity is
expected to be 0.69 for 12 hour runs and 0.82 for 6 hour runs. Thus the OLAP background could easily be up to
two times smaller than that calculated in Table \ref{tb:sigmas}, even
at high luminosity running.
\begin{figure}[t]
\centering
\mbox{
\subfigure[]{\includegraphics[width=0.5\textwidth,height=5.85cm]{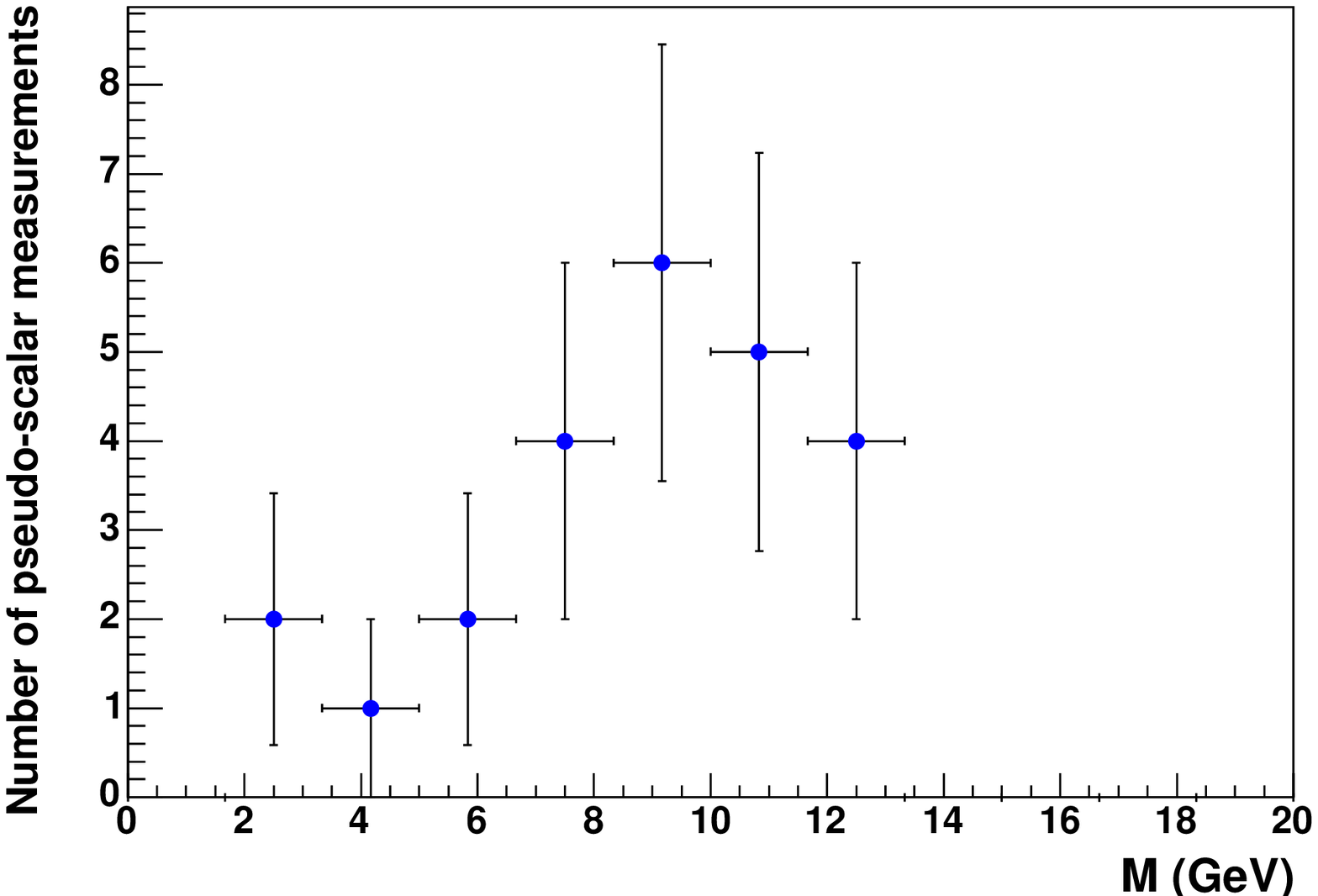}}
        \subfigure[]{\includegraphics[width=0.5\textwidth]{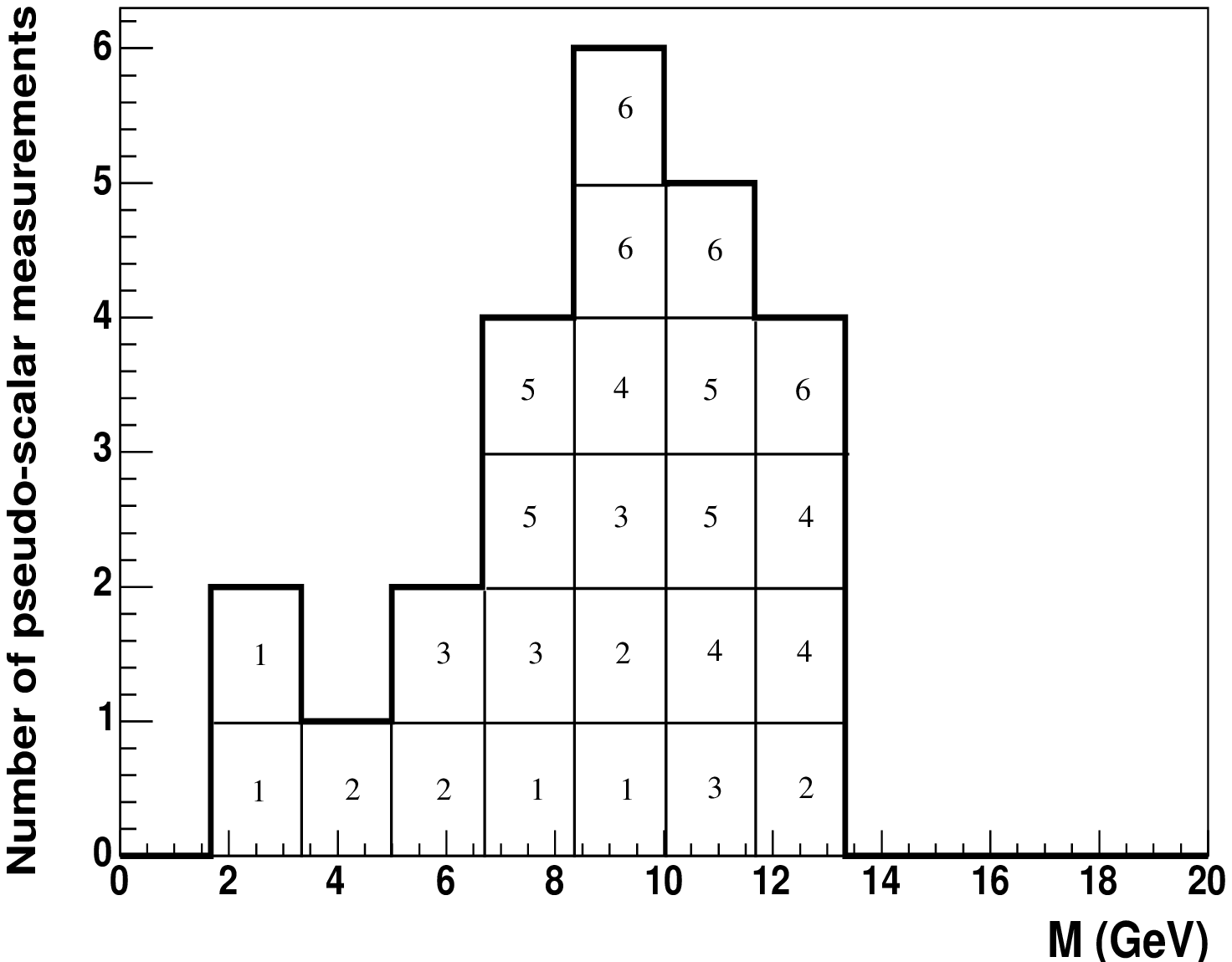}}
}
\caption{(a) A typical $a$ mass measurement. (b) The same content as (a) but with the breakdown showing
the 4 Higgs mass measurements for each of the 6 events, labelled $1-6$ in the histogram. }
\label{fig:amassdata}
\end{figure}

To conclude, we show in Figure~\ref{fig:amassdata}(a) a typical distribution in the measured mass of the light pseudo-scalar $a$.
Since we make 4 mass measurements per event, the 6 signal events each contribute 4 entries in the histogram.
Figure \ref{fig:amassdata}(b) shows the distribution of masses obtained for each event; the integer in each box labels one of the 6 signal events.  The figure is indicative of what  one could obtain with 180 fb$^{-1}$ of data collected at $3 \times 10^{33}$ cm$^{-2}$s$^{-1}$. By considering many pseudo-data sets, we conclude that a typical experiment would yield $m_a = 9.3 \pm 2.3$ GeV, which is in re-assuringly good agreement with the expected value of 9.7 GeV.

\section{Conclusions}
\label{sec:conclusions}

Both the Standard Model and Minimal Supersymmetric Model have theoretical problems with regard to naturalness, hierarchy
and fine-tuning, especially after including the LEP constraints on the
lightest CP-even Higgs boson. Models which avoid these problems are
typified by unusual decay modes for the lightest of the CP-even Higgs
bosons. As a result, there is no guarantee that LHC Higgs discovery
will be possible using the conventional modes explored for the SM and
MSSM. A particularly attractive model that avoids the theoretical
problems of and experimental constraints on the SM and MSSM is the
Next-to-Minimal Supersymmetric SM. Hierarchy and fine-tuning issues are
absent provided the soft-SUSY-breaking parameters are chosen in such a
way that sparticles are relatively light, leading to a light CP-even
$h$ with $m_h\sim 100\gev$ and SM-like couplings to gauge bosons and
fermions.  The $h$ of the NMSSM must have escaped LEP detection by
virtue of $h\to aa\to 4\tau$ or $4j$, with the former being the most
natural result in the context of the theory.  We have demonstrated
that the prospects of utilizing CEP to observe such an NMSSM $h$
decaying via $h\to aa\to 4 \tau$, are good, provided that sufficient
integrated luminosity is available. The viability of the $pp\to
p+h+p$ detection channel is especially important given that no other LHC
modes have as of yet been shown to be viable and given the possibility to
extract the mass of the $h$ and $a$ on an event-by-event basis.

With regard to the need for high instantaneous luminosity, we have demonstrated
that the analysis can work even at high instantaneous LHC
luminosity, with the optimum being about ${\cal L}\sim 5\times
10^{33}~{\rm cm}^{-2}~\rm{sec}^{-1}$.  This is because of our ability
to bring pile-up under control, which in turn is a consequence of the fact
that our analysis makes very little use of the hadronic and
electromagnetic calorimeters and instead relies more on tracking and
the muon detectors.  

Although we have not explored the matter in this paper, it seems quite
likely that such a track-based analysis could also be utilized in
other CEP processes, for example \linebreak $pp \to p+h+p$ with $h \to \tau^+
\tau^-$. A heavily track-based approach could also provide an
interesting analysis route for more mainstream LHC processes in order
to help diminish the influence of pile-up, {\it e.g.} in analyses
which aim to exploit the presence of rapidity gaps. An obvious example
of such a process is Higgs production via weak boson fusion (WBF).
WBF is traditionally distinguished from Higgs production via gluon
fusion (and backgrounds) by requiring the presence of two highly
energetic jets at large absolute rapidity values (recoiling from the
emitted virtual $W$'s). Since the virtual $W$'s are color singlets,
the rapidity region between the recoil jets and the centrally produced
Higgs decay products should have reduced jet activity compared to
backgrounds and other Higgs production processes.  An analysis similar
to that presented here could be used to select events according to the
numbers of tracks they contain in the central region. This approach would be particularly
interesting for $H \to \tau^+ \tau^-$ (in the SM or otherwise) where one $\tau$ decays to a
muon. The muon then defines, very precisely, the location of the
primary vertex and one could cut on the number of additional tracks
pointing back to that vertex. It will also be important to explore
whether an analysis along these lines might even make viable the
detection of the NMSSM $h\to aa\to 4\tau$ Higgs decay mode when the
$h$ is produced via WBF.  Whatever the primary Higgs decay mode, but
especially in the $h\to aa \to 4\tau$ case, observations of the same
Higgs final state via WBF would be very important for confirming any
Higgs signal seen in CEP.

Finally, we return to the fact that $m_a<2m_\tau$ remains possible in
the NMSSM, although this requires a little more parameter tuning.
Thus, there is a need to identify techniques that might yield a viable
signal for the resulting $h\to aa\to 4j$ final state. Similarly we
should keep in mind that $m_h > 110$ GeV with large branching ratios
for $h \to aa$ and $a \to b\bar{b}$ is not excluded by LEP.  It is
therefore also important to establish the viability of observing this
channel in CEP.

\section*{Acknowledgements}
Thanks to Max Chertok, Thorsten Wengler, Brian Cox, Valery Khoze,
Attila Krasznahorkay, Christian Schwanenberger, Tim Steltzer, Peter Bussey and Bob
McElrath for help and useful discussions. JFG is supported by the
U.S. Department of Energy under grant No. DE-FG03-91ER40674.
This research was also partially supported by the U.K.'s STFC.


\begin{thebibliography}{99}
\bibitem{bechtlepc}  This is the result obtained by P. Bechtle after
running several possible scenarios through the full LEP Higgs
analysis.
\end{thebibliography}
\end{document}